\documentclass[prb,twocolumn,showpacs]{revtex4}
\usepackage{graphicx,epsf}
\usepackage{amsfonts}
\def\bn{{\bf n}}
\def\br{{\bf r}}
\def\bk{{\bf k}}
\def\bL{{\bf L}}

\begin{document}

\newcount\bozza \bozza=0
\ifnum\bozza=1
\newdimen\shift \shift=-2truecm
\def\lb#1{
{\label{#1}\rlap{\kern\shift{$\scriptstyle#1$}}}}
\else\def\lb#1{\label{#1}} \fi

\def\insertplot#1#2#3#4{\begin{minipage}{#2}
\vbox {\hbox to #1 {\vbox to #2 {\vfil \includegraphics{#4.ps}#3}}}
\end{minipage}}

\title{Dissipative dynamics of topological defects in frustrated Heisenberg
spin systems}
\author{V.\ Juricic$^{1,2}$, L.\ Benfatto$^1$, A.\ O.\
Caldeira$^3$, and C.\ Morais Smith$^{1,2}$}

\affiliation{$^1$D\'epartement de Physique, Universit\'e de
Fribourg, P\'erolles, CH-1700 Fribourg, Switzerland.\\
$^2$ Institute for Theoretical Physics, University of Utrecht,
Leuvenlaan 4, 3584 CE Utrecht, The Netherlands.\\ $^3$ Instituto
de F\'\i sica Gleb Wataghin, Universidade Estadual de Campinas,
13083-970, Campinas, SP, Brazil}

\begin{abstract}

We study the dynamics of topological defects of a frustrated spin
system displaying spiral order. As a starting point we consider
the $SO(3)$ nonlinear sigma model to describe long-wavelength
fluctuations around the noncollinear spiral state. Besides the
usual spin-wave magnetic excitations, the model allows for
topologically non-trivial static solutions of the equations of
motion, associated with the change of chirality (clockwise or
counterclockwise) of the spiral. We consider two types of these
topological defects, single vortices and vortex-antivortex pairs,
and quantize the corresponding solutions by generalizing the
semiclassical approach to a non-Abelian field theory. The use of
the collective coordinates allows us to represent the defect as a
particle coupled to a bath of harmonic oscillators, which can be
integrated out employing the Feynman-Vernon path-integral
formalism. The resulting effective action for the defect indicates
that its motion is damped due to the scattering by the magnons. We
derive a general expression for the damping coefficient of the
defect, and evaluate its temperature dependence in both cases, for
a single vortex and for a vortex-antivortex pair. Finally, we
consider an application of the model for cuprates, where a spiral
state has been argued to be realized in the spin-glass regime. By
assuming that the defect motion contributes to the dissipative
dynamics of the charges, we can compare our results with the
measured inverse mobility in a wide range of temperature.  The
relatively good agreement between our calculations and the
experiments confirms the possible relevance of an incommensurate
spiral order for lightly doped cuprates.

\end{abstract}
\maketitle

\section{Introduction}

Two-dimensional frustrated Heisenberg spin systems with
noncollinear or canted order have attracted much attention
recently. Noncollinear order arises due to frustration, which may
originate from different sources.  The most common kind of
frustration is realized in antiferromagnets on a two-dimensional
(or three-dimensional stacked) triangular lattice. Prototypes of
these geometrically frustrated magnets are
pyrochlores.\cite{kawamura,pel,del1} A second source of
frustration may be a competition between nearest-neighbor and
further-neighbor exchange interactions between spins. Typical
examples are helimagnets, where a magnetic spiral is formed along
a certain direction of the lattice.\cite{kawamura} A third kind of
frustration may occur by chemical doping of a magnetically ordered
system. In this case, the spin-current of the itinerant doped
charges couples to the local magnetic moment of the magnetic host,
leading to the formation of a noncollinear magnetic state. This
situation may be realized in lightly doped cuprate
superconductors. \cite{siggia881,siggia882,
siggia89,kane90,schulz,weng,mori,nils,juricic}

The main characteristic of the noncollinear state is that the spin
configuration must be described by a set of three orthonormal
vectors or, alternatively, by a rotational matrix which defines
the orientation of this set with respect to some fixed reference
frame. As a consequence, the order-parameter space is isomorphic
to the three-dimensional rotational group $SO(3)$, and in the
low-temperature phase, when the rotational symmetry is fully
broken, three spin-wave modes are present in the system, instead
of two, as in the nonfrustrated case.  Moreover, topological
defects may arise in the system, associated with the chiral
degeneracy of the spiral, which can rotate clockwise or
counterclockwise. Because the order parameter space has a
nontrivial first homotopy group, $ \pi_1[SO(3)]=\mathbb{Z}_2$, the
topological excitations are vortex-like. On the other hand,
skyrmions are not present because the second homotopy group of the
$SO(3)$ is trivial, $ \pi_2[SO(3)]=0$. \cite{dombre}

A convenient field-theoretical description of frustrated
Heisenberg systems in the long-wavelength limit is provided by the
$SO(3)$ nonlinear sigma (NL$\sigma$) model.
\cite{dombre,apel,delamotte,klee,azaria}  Its critical behavior in
two dimensions has been extensively investigated, both in the
absence and in the presence of topological excitations. Studies in
the former case have revealed  a dynamical enhancement of the
symmetry from $O(3)\otimes O(2)$ to $O(4)$ under renormalization
group flow in ${d}=2+\epsilon$, which means that in the critical
region all the three spin-wave modes have the same
velocity.\cite{apel,azaria} When topological excitations are
included,  a complex finite-temperature behavior is
found.\cite{wintel1} Numerical studies, as well as analysis
involving entropy and free energy arguments, indicate the
occurrence of a transition driven by vortex-antivortex pairs
unbinding at a finite temperature
$T_v$.\cite{kawamura1,southern,wintel2,caffarel} In contrast to
the $XY$ case, here vortices and spin-waves are coupled already in
the harmonic approximation, and anharmonic spin-wave interactions
yield a finite correlation length for arbitrarily low
temperatures. \cite{polyakov,azaria} Therefore, the transition
mediated by vortices is rather a crossover than a true
Kosterlitz-Thouless (KT) transition.\cite{kosterlitz} Free
vortices start to proliferate at the temperature $T_v$, similarly
as vortices in the $XY$ model do above the KT-transition
temperature.

In the present paper we study the physical properties of
frustrated Heisenberg spin system, which are sensitive to the
dynamics of the above-mentioned topological defects.  The approach
we use has been employed to describe the dynamics of excitations
in a very broad class of one- or two-dimensional
systems.\cite{amir} The central idea is the application of the
collective-coordinate method \cite{raja} to quantize a non-trivial
static solution of the classical equation of motion of the
field-theoretical model in question. In our case, we find that
single vortex-like excitations or vortex-antivortex pairs are the
localized static solution of the $SO(3)$ NL$\sigma$ model. A
proper description of the quantum levels associated with these
solutions is provided, on the semi-classical level, by a theory in
which the topological excitation is represented by a single
quantum mechanical variable coupled to a bath of quantum harmonic
oscillators, which are the fluctuations about the classical
solution itself. Thus, the resulting effective model represents a
particle (the topological defect) scattered by the linearized
excitations of the system. The latter can be integrated out using
the standard system-plus-reservoir approach, \cite{amir} leading
to a dissipative equation of motion for the topological
excitation. As a consequence, any physical property of these
systems that depends on the motion of the topological excitations
may be expressed in terms of transport coefficients - such as
mobility and diffusion - of these damped defects. Since we do
neglect any interaction between the defects, our results are only
valid for a diluted gas of topological excitations. Part of our
results concerning the mobility of a vortex-antivortex defect has
been recently  published in Ref.\ \onlinecite{juricic}. Here,
besides a complete presentation of the technical details, we
discuss also the transport at finite frequencies and we compare
-qualitatively and quantitatively- the cases where the defect is
represented by a single vortex or by a vortex-antivortex pair.
Moreover, in the light of new experimental results by Ando {\it et
al.}, \cite{ando1} we discuss the relevance of the single vortices
for the transport in the spin-glass phase of cuprates.

The structure of the paper is the following. Starting from  the
$SO(3)$ NL$\sigma$ model, we derive in Sec.\ II  the quantum
Hamiltonian describing the dynamics of the topological defect
coupled to a bath of magnetic excitations. In Sec.\ III the
equation governing the evolution of the reduced density matrix for
the topological defect is obtained and the influence functional,
which describes the effect of the magnon bath on the dynamics of
the vortices is evaluated. Section IV is devoted to the derivation
of the  effective action for the defect after the magnons have
been integrated out, and in Sec.\ V the inverse mobility is
calculated. In Sec.\ VI we discuss  how a spiral state may be
realized in cuprates and we then apply our results to this
specific case. Section VII contains our conclusions. Details of
the calculations are given in the Appendices.

\section{The Model}

In the spiral state the spin configuration $\bf S$ at each site
$\br$ is described by means of a dreibein order parameter $n^a_k
\in SO(3)$ with $k = 1,2,3$ and $n_k^a n_q^a = \delta_{kq}$,
\cite{apel} so that
\begin{equation}
\lb{incomm}
\frac{{\bf S}}{S}=\bn=\bn_1 \cos(\bk_s\cdot \br)-\bn_2
\sin(\bk_s\cdot \br),
\end{equation}
where $S = |{\bf S}|$ and the wave vector ${\bf k}_s= (\pi/a
,\pi/a)+ {\bf Q}$, with $a$ denoting the lattice constant. Here
${\bf Q}=(2\pi/m_x a,2\pi/ m_y a)$ measures the {\em
incommensurate} spin correlations . Indeed, the magnetic
susceptibility corresponding to the spin modulation (\ref{incomm})
has two peaks at ${\bf k}_s$ and $-{\bf k}_s$ (equivalent to
$(\pi/a ,\pi/a)- {\bf Q}$), as represented in Fig.\ 1 in the case
of $m_x=-m_y$. The resulting spin order for ${\bf n}_1$ and ${\bf
n}_2$ in the plane is represented in Fig.\ 2, where $m_x=-m_y=20$.
Observe that the periodicity of the spin texture is $2\pi/{\bf Q}$
for even values of $m_x,m_y$, and twice it for odd values.

\begin{figure}[htb]
\begin{center}
\includegraphics[width=5cm,angle=0]{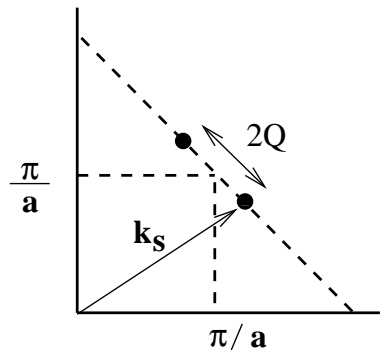}
\end{center}
\caption{Incommensurate magnetic response for the spiral spin
modulation (\ref{incomm}). The magnetic susceptibility
corresponding to the spiral order with the wave vector ${\bf k}_s$
exhibits two peaks at the points $(\pi/a,\pi/a) \pm {\bf Q}$
marked by a circle in the figure. In this case ${\bf Q}$ has
finite components in both the $x$ and $y$ directions, and the
distance between the peaks is twice the modulus of ${\bf Q}$.}
\end{figure}

\begin{figure}[htb]
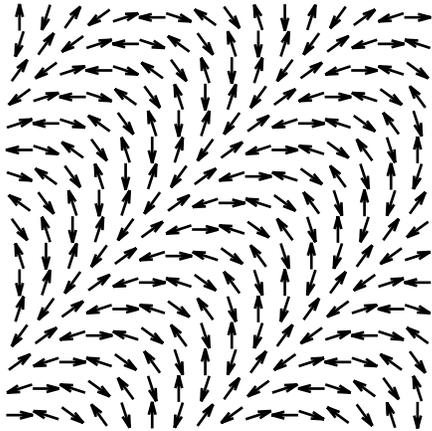

\begin{center}
\insertplot{160pt}{160pt}{ }{spiral}
\end{center}
\caption{Spin background corresponding to Eq.~(\ref{incomm}) and to the
case depicted in Fig.\ 1. Here a value of ${\bf Q}=(2\pi/20a, -2\pi/20a)$
has been choosen.}
\end{figure}

As discussed in Refs.\ \onlinecite{apel,delamotte,klee}, a proper
continuum field theory for the spiral state is provided by the
$SO(3)$ quantum NL$\sigma$ model,
\begin{eqnarray}
\lb{eqkm} {\cal S}=\int dt d^2{\bf x}  [\kappa_k(\partial_t
\bn_k)^2- p_{k\alpha} \left(\partial_\alpha \bn_k\right)^2].
\nonumber
\end{eqnarray}
Here, the index $\alpha$ stands for the spatial coordinates and
summation over repeated indices is understood. The spatial
anisotropy of the spin stiffness $p_{k\alpha}$ depends on the
components $Q_\alpha$ of the incommensurate wave vector. Since at
the fixed point all the spin-wave velocities are equal,
\cite{apel,azaria} we will consider the case
$\kappa_k\equiv\kappa$, $p_{k\alpha}\equiv p_\alpha$ and we will
choose a system of coordinates parallel $(x_\parallel)$ and
perpendicular $(x_\perp)$ to the spiral axis, respectively,

\begin{equation}
\lb{eqkm1}
\mathcal{S}=\kappa\int dt
dx_\perp dx_\parallel[(\partial_t\bn_k)^2-
c_\perp^2(\partial_\perp \bn_k)^2-c_\parallel^2
(\partial_\parallel \bn_k)^2],
\end{equation}
where $c_{\perp}\equiv\sqrt{p_\perp/\kappa}$ and
$c_{\parallel}\equiv\sqrt{p_\parallel/\kappa}$ are the spin wave
velocities perpendicular and parallel to the spiral axis. Even
though for the moment we will keep our derivation on general
grounds, in Sec.\ VI we will specify the values of the parameters
$\kappa$ and $c$ for the case of cuprates, where they can be
related to measurable quantities.

Given the action (\ref{eqkm1}) as our starting model, our first aim is to
analyze whether the equations of motion admit topologically non-trivial
solutions. For that purpose, it is convenient to introduce an equivalent
representation of the $n^a_k$ order parameter through an element $g \in
SU(2)$ as $n^a_k=(1/2) {\rm tr}[\sigma^a g \sigma^k g^{-1}]$, where
$\sigma^a$ are Pauli matrices, and to introduce the fields
\begin{equation}\lb{A}
A_\mu^a=\frac{1}{2i}
{\rm tr}[\sigma^a g^{-1} \partial_\mu g],
\end{equation}
which are related to the first derivatives of the order parameter through
$\partial_\mu n^a_k=2\epsilon_{ijk}A^i_\mu n^a_j$.  \cite{polyakov} Here,
$\partial_\mu\equiv (\partial_t,\nabla)$ and
$\epsilon_{123}=\epsilon^{123}=1$. Using that $(\partial_\mu \bn)^2= 8{\bf
A}_\mu^2$ (no summation over index $\mu$ is imposed here), the action
(\ref{eqkm1}) reads
\begin{equation}\lb{action}
\mathcal{S} = 8\kappa \int dt
dx_\parallel dx_\perp
\left({\bf A}_0^2-c_\perp^2{\bf A}_{\perp}^2-c_\parallel^2
{\bf A}_\parallel^2\right).
\end{equation}

The above action  may be mapped to an isotropic form by
introducing the coordinates ${\bf r}=(x, y)$ with
$$x=\sqrt{\frac{c_\parallel}{c_\perp}}x_\perp, \qquad
y=\sqrt{\frac{c_\perp}{c_\parallel}}x_\parallel.$$
We then find
\begin{equation}
\lb {eq:isoaction}
\mathcal{S} = \mathcal{N} \int dt d^2{\bf r} {\bf A}_\mu^2
 \equiv \mathcal{N}\int dt d^2{\bf r}
\left({\bf A}_0^2-c^2{\bf A}_\alpha^2\right)
\end{equation}
with the isotropic spin-wave velocity
$c=\sqrt{c_{\parallel}c_{\perp}}$ and the constant
$\mathcal{N}=8 \kappa$. The most generic expression for
the element $g$ is given by
\begin{equation}
\lb{gen}
g[\vec{\alpha}]=\exp\left(\frac{i}{2} {\vec \alpha}({\bf
 r},t)\cdot {\vec \sigma} \right),
\end{equation}
and the corresponding Lagrangian obtained from the action (\ref{eq:isoaction})
reads
\begin{equation}
\lb{lgen}
L_0 = \frac{1}{4}\mathcal{N} \int d^2\br(\partial_\mu \vec \alpha)^2,
\end{equation}
with $\partial_\mu A\partial_\mu B\equiv \partial_tA \partial_tB-c^2\nabla
A\nabla B$. By making the ansatz \cite{wintel1}  $\vec\alpha({\bf
r},t)=\vec m \Psi({\bf r},t)$, where $\vec m$ is a constant unit vector and
$\Psi$ a scalar function, the Lagrangian (\ref{lgen}) reduces to
\begin{equation}
\lb{tdfree}
L_0 = \frac{1}{4}\mathcal{N} \int d^2\br(\partial_\mu \Psi)^2.
\end{equation}
The equation of motion for the field $\Psi$,
\begin{equation}\lb{tdeq}
\partial_t^2\Psi-c^2\nabla^2\Psi=0,
\end{equation}
possesses static topologically non-trivial solutions in the form
of a single-vortex defect at ${\bf R}=(X,Y)$,
\begin{equation}
\lb{1v}
\Psi_{1v}=\arctan\left(\frac{x-X}{y-Y}\right),
\end{equation}
and a vortex-antivortex pair
\begin{eqnarray}
\Psi_{2v}&=&\arctan\left(\frac{x-X_1}{y-Y_1}\right)-
\arctan\left(\frac{x-X_2}{y-Y_2}\right)\nonumber \\ &=&\arctan
\lb{2v}
\left\{\frac{[{\bf d}\times({\bf r}-{\bf R})]_{z}}{[({\bf r} - {\bf R})^2-
d^2/4 ]}\right\},
\end{eqnarray}
where now ${\bf R} = ({\bf R}_1 + {\bf R}_2)/2$ is the center of
mass and ${\bf d}={\bf R}_2 - {\bf R}_1$ the relative coordinate
of the defect pair. If in Eqs.~(\ref{1v}) and (\ref{2v}) the role
of $x$ and $y$ coordinates is interchanged, one only changes the
vorticity.  Without loss of generality, we may assume the unity
vector to be in the $z$-direction, ${\bf m}=\hat{e}_z$. Thus, the
${\bf n}_k$ fields which define the spin configuration according
to Eq.~(\ref{incomm}), are given by ${\bf n}_1=(\cos \Psi,
-\sin\Psi, 0), {\bf n}_2=(\sin\Psi,\cos \Psi, 0), {\bf n}_3=(0, 0,
1 )$. The spin patterns corresponding to a single vortex ($\Psi$
from Eq.~(\ref{1v})) or to a vortex-antivortex pair ($\Psi$ from
Eq.~(\ref{2v})) are represented in Fig.\ 3 and Fig.\ 4,
respectively.

\begin{figure}[htb]
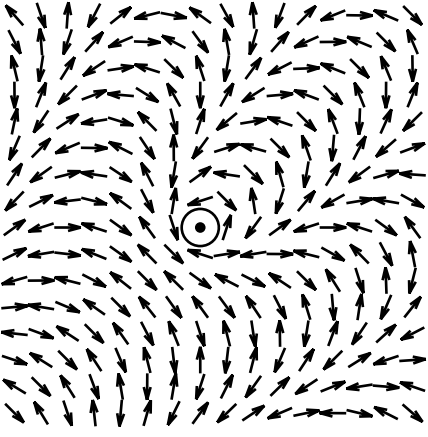

\begin{center}
\insertplot{160pt}{160pt}{ }{one_vortex}
\end{center}
\caption{Spin background corresponding to the one-vortex solution of
Eq.~(\ref{1v}). The center of the vortex is marked by the circle. The
spiral incommensurability ${\bf Q}$ is the same used in Fig.\ 2.}
\end{figure}

\begin{figure}[htb]
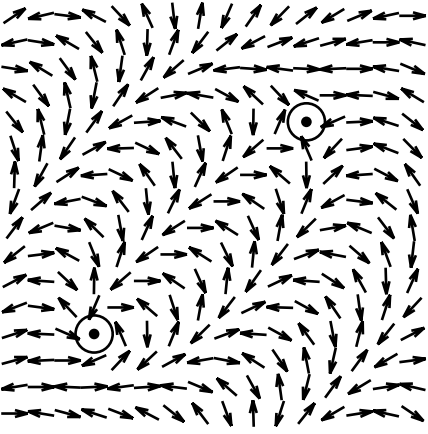

\begin{center}
\insertplot{160pt}{160pt}{ }{two_vortex}
\end{center}
\caption{Spin background corresponding to the vortex-antivortex solution of
Eq.~(\ref{2v}). The centers of the vortices are marked by a circle. The
spiral incommensurability ${\bf Q}$ is the same used in Fig.\ 2.}
\end{figure}

The main difference between the two possible static solutions
(\ref{1v}) and (\ref{2v}) is their energy. As shown in App.\ A,
the energy of a single-vortex diverges with the logarithm of the
system size $\ell$, $E[\Psi_{1v}]\propto \ln \ell$. On the other
hand, the vortex-antivortex pairs have finite energy, depending on
the distance $d$ between defects, $E[\Psi_{2v}]\propto \ln d$. A
similar situation is realized in the standard $XY$ model, where
indeed the presence of single defects below the KT transition is
not energetically favorable in the thermodynamic
limit.\cite{kosterlitz} However, in the case of the model
(\ref{eqkm1}), which possesses asymptotic freedom, the correlation
length $\xi$ is finite at any finite
temperature,\cite{azaria,polyakov} so that the logarithmic
divergence of the single-vortex energy should be understood up to
the length scale $\xi$, $E[\Psi_{1v}]\propto \ln \xi$. In
addition, the energy of the vortex-antivortex pair should be
bounded below at distances of the order of few lattice spacings,
which is the intrinsic cutoff of the theory. Because the procedure
that we describe in the following does not depend on the exact
form of the static solution, we will refer to a static topological
defect solution $\Psi_v$ specifying only at the end of the
calculations the differences between the cases (\ref{1v}) and
(\ref{2v}).

Following a procedure analogous to the one described  in Ref.\
\onlinecite{raja} to quantize  the kink solution of the scalar
field theory, we analyze now the effect of the fluctuations around
the static topologically nontrivial configuration, which is a
saddle point of the action that corresponds to the Lagrangian
(\ref{tdfree}). In order to reach this aim, we write the generic
field $g\in SU(2)$ of Eq.\ (\ref{gen}) in the form of a product of
the field $g_s$ corresponding to a static solution $\vec m
\Psi_v({\bf r})$ and the field $g_\varepsilon$ corresponding to
the fluctuations around it,
\begin{equation}
\lb{gfull}
g({\bf r}, t)=g_s\left[\Psi_v({\bf r})\right]g_\varepsilon
\left[{\vec \varepsilon}({\bf r}, t)\right],
\end{equation}
where
$$
g_\varepsilon[{\vec\varepsilon}]=\exp\left(\frac{i}{2}
{\vec \varepsilon}\cdot{\vec \sigma}\right).
$$
Observe that the description of the fluctuations via  Eq.\
(\ref{gfull}) differs from the standard approach used for a scalar
field theory, \cite{raja} and it is related to the symmetry
properties of the order parameter. Indeed, since the full $g$ has
to be an element of the $SU(2)$ group, and both $g_s,g \in SU
(2)$, then the fluctuations $g_\varepsilon$ around $g_s$ have to
belong to $SU(2)$ as well. If instead, we had used the expansion
$\vec \alpha=\vec m\Psi_v+\vec \varepsilon$, the equations of
motion for the $\vec \varepsilon$ field would have been
independent of the static solution $\Psi_v$, leading to a failure
of the semiclassical expansion.  Using Eqs.\ (\ref {A}) and (\ref
{gfull}), we can express the action (\ref {eq:isoaction}) in terms
of the fields $\Psi_v$ and ${\vec \varepsilon}$.  Retaining only
terms up to second order in ${\vec \varepsilon}$, we find
(Appendix B)
\begin{eqnarray}
A^a_\mu &=& \frac{1}{2}m^a\partial_\mu\Psi_v\left(1-\frac
{\bar{\varepsilon}^2+\varepsilon^2_z}{2}\right)+\frac{1}{4}
\varepsilon^a
\varepsilon^b m^b \partial_\mu\Psi_v \nonumber \\
&+&\frac{1}{2}\epsilon^{abc}\varepsilon^bm^c\partial_\mu
\Psi_v
+\frac{1}{2}\partial_\mu\varepsilon^a+\frac{1}{4}\epsilon
^{abc}\varepsilon^b\partial_\mu\varepsilon^c,\nonumber
\end{eqnarray}
where $\bar{\varepsilon}^2=\varepsilon_x^2+\varepsilon_y^2$.
The corresponding Lagrangian then reads (App. B)
\begin{equation}
\lb{totl}
L = L_0 + \mathcal{N} \int  d^2{\bf r} \mathcal{L}_1
\end{equation}
with $L_0$ given by Eq.\ (\ref{tdfree}) and
\begin{eqnarray}
\mathcal{L}_1[\Psi_{v}]&=&\frac{1}{4}(\partial_\mu{\vec \varepsilon})^2
+\frac{1}{2}({\vec m}\cdot \partial_\mu{\vec
 \varepsilon})\partial_\mu\Psi_{v}\nonumber \\
&+& \frac{1}{4}\epsilon^{abc}\partial_\mu\varepsilon^a
\varepsilon^b m^c \partial_\mu\Psi_{v}=\nonumber\\
&=&\frac{1}{4}(\partial_\mu
\bar{\varepsilon})^2
+\frac{1}{4}\bar{\varepsilon}^2(\partial_\mu\theta)^2+\frac
{1}{4}(\partial_\mu\varepsilon^z)^2
\nonumber\\
 \lb{L2v1}
&+&\frac{1}{2}\partial_\mu\varepsilon^z\partial_\mu
\Psi_{v}-\frac{1}{4}\bar{\varepsilon}^2
\partial_\mu\Psi
_{v}\partial_\mu\theta.
\end{eqnarray}
Here, we used the fact that ${\bf m}=\hat{e}_z$ and  introduced
polar coordinates $\vec \varepsilon=(\bar{\varepsilon} \cos\theta,
\bar{\varepsilon}\sin\theta,\varepsilon_z)$. Since the Lagrangian
$\mathcal{L}_1$ is evaluated at the vortex-like solution $\Psi_v$
of Eq.\ (\ref{tdeq}), the equations of motion for the fluctuations
around the topological defect also depend on $\Psi_v$
\begin{equation}\lb{flepsilon}
\bar{\varepsilon}:\quad (\partial_t^2-c^2\nabla^2)
\bar{\varepsilon}-
\bar{\varepsilon}(\partial_\mu\theta)^2-c^2
\bar{\varepsilon}\nabla\Psi_{v}
\nabla\theta=0,
\end{equation}
\begin{equation}\lb{theta}
\theta:\qquad\qquad \partial_\mu(\bar{\varepsilon}^2\partial_\mu\theta)
+\frac{c^2}{2}\nabla(\bar{\varepsilon}^2\nabla\Psi_{v})=0,
\end{equation}
\begin{equation}\lb{epsilonz}
\varepsilon_z:\qquad\qquad (\partial^2_t-c^2\nabla^2)
\varepsilon^z=0.
\end{equation}
Eq.\ (\ref{theta}) admits the solution $\theta=\Psi_{v}/2$,
whereas Eq.\ (\ref{epsilonz}) indicates that the field
$\varepsilon_z$ is free. By using these two conditions, we can
rewrite the total Lagrangian ${L}$ in Eq.\ (\ref{totl}) as
\begin{equation}
\lb{cc3}
L=\frac{{\cal N}}{4}\int d^2{\bf r}
\left[(\partial_\mu\Psi_{v})^2+(\partial_\mu
\bar{\varepsilon})^2-\frac{1}{4}\bar{\varepsilon}^2(\partial_\mu\Psi_{v})^2\right],
\end{equation}
and the equation of motion (\ref{flepsilon}) as
$$
\left[\partial^2_t-c^2\nabla^2-\frac{1}{4}(\nabla\Psi_{v}
)^2\right]\bar{\varepsilon}=0.
$$
Since the field $\Psi_v$ in the previous equation does not depend
on time, we decompose the field $\bar{\varepsilon}$ into its time-
and space-dependent parts
\begin{equation}\lb{expansion}
\bar{\varepsilon}({\bf r},t)=\sum_{nm}  q_{nm}(t)\eta_{nm}({\bf
r}),
\end{equation}
and
identify the normal modes $\eta_{nm}$ with the eigenfunctions of
the operator
\begin{equation}\lb{schroedinger}
c^2\left[\nabla^2+V({\bf r})\right]\eta_{nm}=-\omega^2_{nm}\eta_{nm}.
\end{equation}
This equation has the typical form of a Schr\" odinger-like
equation for a particle scattered by a potential $V({\bf
r})=(\nabla\Psi_{v})^2/4$. The two indices $n$ and $m$ refer,
respectively, to the radial and angular part of the wavefunction.
By using a standard approach to scattering problems in two
dimensions one may express the wavefunctions $\eta_{nm}$ in terms
of the eigenfunctions of the free problem ($V=0$), corrected by a
phase shift $\delta_m$ due to the scattering by the potential
$V(\bf r)$, \cite{morse}
\begin{equation}\lb{eta}
\eta_{nm}=\frac{1}{2}\sqrt{\frac{k_{nm}}{2\ell}}e^{im\vartheta}
\left[H^{(1)}_{|m|}(k_{nm}r)+e^{-2i\delta_m}H^{(2)}_{|m|
}(k_{nm}r)\right].
\end{equation}
Here, $H^{(1,2)}_{|m|}$ are Hankel functions of the first and
second kinds, $m$ is an integer, and $\vartheta$ is a polar angle.
The $k_{nm}$ values are determined by requiring the vanishing of
the wavefunction (\ref{eta}) at the boundary $r=\ell$. By using
the asymptotic form of the Hankel functions we obtain
$k_{nm}\ell=(2n+1)\pi/2+(2|m|+1)\pi/4-\delta_m$, where $n$ is a
positive integer. Since the field ${\bar \varepsilon}$ is real, we
may rewrite the expansion (\ref{expansion}) in the form
\begin{equation}\lb{expansion1}
\bar{\varepsilon}({\bf r},t)=\sum_{n,m\geq0} \left(
q_{nm}(t)\eta_{nm}({\bf r})+q^*_{nm}(t)\eta^*_{nm}({\bf r
})\right),
\end{equation}
where we used the identities
$\eta_{n,m}=e^{-2i\delta_m}\eta^*_{n,-m}$ and
$\delta_m=\delta_{-m}$. Note that the sum  in Eq.\
(\ref{expansion1}) is over the positive angular momenta, as  will
be the case in what follows.

The static defect solution $\Psi_v({\bf r})$ of Eq.\ (\ref{tdeq})
is invariant under translation of the center of the defect (i.e.,
the position of the vortex or the center of mass of the
vortex-antivortex pair). A consequence of this invariance
\cite{raja} is that Eq.\ (\ref{schroedinger}) admits
zero-frequency modes.  A consistent treatment of them requires the
use of the collective coordinate method.\cite{raja,amir} The
center of mass of the defect is then promoted to a dynamical
variable, yielding
\begin{equation}\lb{cc1}
\Psi_{v}({\bf r})\rightarrow\Psi_{v}({\bf r}-{\bf R}(t)),
\end{equation}
and
\begin{equation}\lb{cc2}
\bar{\varepsilon}({\bf r},t)\rightarrow \bar{\varepsilon}({\bf r}-
{\bf R}(t),t)\equiv\sum_{nm}(q_{nm}(t)\eta_{nm}({\bf r}- {\bf
R}(t))+C.c.),
\end{equation}
where the last sum is over all non-zero-frequency modes.  By inserting
these expressions into the full Lagrangian (\ref{cc3}) evaluated at the
saddle-point solution we obtain
\begin{eqnarray}\lb{dotpsi}
\frac{\mathcal{N}}{4}\int dt d^2{\bf r} (\partial_t\Psi_{v})^2&=&
\frac{\mathcal{N}}{4}\int dt d^2{\bf r}\dot{R}_\alpha\dot{R}_\beta
\partial_\alpha\Psi_{v}\partial_\beta\Psi_{v}\nonumber\\
&=& \frac{M}{2}\int dt \dot{{\bf R}}^2(t),
\end{eqnarray}
where
$$
M=\frac{\mathcal{N}}{2}\int d^2{\bf r}(\nabla\Psi_{v})^2
$$
is the mass of the topological defect, which is proportional to
its energy (see Appendix A). The time derivative of the field
$\bar{\varepsilon}$ yields
\begin{eqnarray}\lb{doteps}
&&\frac{\mathcal{N}}{4}\int dt d^2{\bf r} (
\partial_t\bar{\varepsilon})^2= \frac{\mathcal{N}}{4}
\sum_{nm,kl}\int dt d^2{\bf r}
\left[\dot{q}^*_{nm}(t)\eta_{nm}^* \right. \nonumber\\
&-& \left. q_{nm}^*
\partial_\alpha\eta^*_{nm}
\dot{R}_\alpha(t)+C.c.\right] \left[\dot{q}_{kl}(t)
\eta_{kl}-q_{kl}\partial_\beta\eta_{kl}
\dot{R}_\beta(t)\right.\nonumber\\
&+&\left. C.c.\right] = \frac{\mathcal{N}}{2}\sum_{nm}\int dt
\left[|\dot{q}_{nm}|^2+\sum_{kl} \dot{{\bf R}}(t)
\left(\dot{q}_{nm}q^*_{kl}{\bf G}^*_{nm,kl}\right.\right.
\nonumber\\
&+&\left.\left. \dot{q}^*_{nm}q_{kl}{\bf G}_{nm,kl}\right)\right],
\end{eqnarray}
where the coupling constants {\bf G} are related to the
eigenfunctions $\eta$ via
$$
{\bf G}_{nm,kl}=\int d^2{\bf r} \eta_{kl}\nabla\eta^*_{nm},
$$
and we neglected terms of order $q^2\dot{\bf R}^2$. Here, we used
that $\int d^2{\bf r}\eta_{nm}\eta_{kl}=0$ and $\int d^2{\bf
r}\eta_{nm}\nabla\eta_{kl}=0$ for $m$ and $l$ positive. By
substituting Eqs.\ (\ref{dotpsi}) and (\ref{doteps}) into the
Lagrangian (\ref{cc3}), we obtain
\begin{eqnarray}
L&=&\frac{1}{2}M\dot{\bf R}^2+\sum_{nm}
(|\dot{q}_{nm}|^2-\omega_{nm}^2|q_{nm}|^2)
\nonumber\\
&+&\sum_{nm,kl} \dot{{\bf R}}(t) \left({\dot q}_{nm}q^*_{kl}{\bf
G}^*_{nm,kl}+ {\dot q}^*_{nm}q_{kl}{\bf G}_{nm,kl}\right),
\nonumber
\end{eqnarray}
where we rescaled $q\rightarrow q\sqrt{\mathcal{N}/2}$. Using that
${\bf G}^*_{nm,kl}=-{\bf G}_{kl,nm}$ and neglecting terms which
are quadratic in ${\bf G}$, the corresponding Hamiltonian reads
\begin{equation}\lb{hclass}
H=\frac{1}{2M}({\bf P}-{\bf P}_E)^2+\sum_{nm}
(|p_{nm}|^2+\omega_{nm}^2|q_{nm}|^2),
\end{equation}
with
$$
{\bf P}_E=\sum_{nm,kl}\left(p_{nm}{\bf G}_{nm,kl}q_{kl}+
p^*_{nm}{\bf G}^*_{nm,kl}q^*_{kl}\right).
$$
Here, ${\bf P}$ is the momentum canonically conjugate to the
center of the defect ${\bf R}$, and $q_{nm}$ and $p_{nm}$ are the
coordinates and momenta of the magnons.  The classical Hamiltonian
(\ref{hclass}) can be promptly quantized by introducing two sets
of independent creation and annihilation operators,
$\hat{a}^\dagger$, $\hat{a}$, and $\hat{b}^\dagger$, $\hat{b}$.
The quantum Hamiltonian reads
\begin{equation}\lb{qham}
\hat{H}=\hat{H}_{v}+\hat{H}_B+\hat{H}_I,
\end{equation}
 where
\begin{equation}\lb{hfree}
\hat{H}_{v}=\frac{\hat{\bf P}^2}{2M}
\end{equation}
is the Hamiltonian of a free defect, and
\begin{equation}\lb{hbath}
\hat{H}_B=\sum_{nm}\hbar\omega_{nm}(\hat{a}^\dagger_{nm}
\hat{a}_{nm}+\hat{b}^\dagger_{nm} \hat{b}_{nm})
\end{equation}
is the Hamiltonian of the bath of magnons which consists of two
independent sets of noninteracting harmonic oscillators described
by the operators ${\hat a}$, ${\hat a}^\dagger$ and ${\hat b}$,
${\hat b}^\dagger$, as it is expected in two dimensions. The
interaction between the bath and the topological defect is
described by the Hamiltonian
\begin{eqnarray}\lb{hint}
\hat{H}_I&=&-\frac{\hbar\hat{{\bf
P}}}{M}\sum_{nm,kl}\left[{\bf D}_{nm,kl}\hat{a}_{nm}
\hat{a}^\dagger_{kl}-{\bf D}_{kl,nm}\hat{b}_{nm}
\hat{b}^\dagger_{kl}\right.\nonumber\\
&+&\left.{\bf C}_{nm,kl}({\hat a}^\dagger_{nm}
\hat{b}^\dagger_{kl}-{\hat a}_{kl}{\hat b}_{nm})\right]
\end{eqnarray}
with the coupling constants given by
\begin{eqnarray}\lb{Dpm}
{\bf D}_{nm,kl}&=&\frac{i}{2}\frac{\omega_{nm}+\omega_
{kl}}{\sqrt{\omega_{nm}\omega_{kl}}}{\bf G}_{kl,nm}
\nonumber\\
{\bf C}_{nm,kl}&=&\frac{i}{2}\frac{\omega_{nm}-\omega_
{kl}}{\sqrt{\omega_{nm}\omega_{kl}}}{\bf G}_{nm,kl}.
\end{eqnarray}
The terms with the coupling constants ${\bf D}/{\bf C}$ describe
the scattering/creation (annihilation) of magnetic excitations by
the defect. Since we consider here only the low-energy dynamics of
the topological defect, we neglect the off-diagonal terms in the
interaction Hamiltonian, i.e., we set ${\bf C}=0$. In the
following we shall integrate out the bath degrees of freedom in
order to study the effective dynamics of the defects. For that
purpose we employ the Feynman-Vernon formalism.

\section{Reduced Density Matrix}
In this section we derive the reduced density matrix for the
defect. The system under consideration consists of two subsystems:
the topological defect and the bath of magnons. Thus, the Hilbert
space of the full system, $\mathcal{H}$, is a direct product of
the subsystem Hilbert spaces
$\mathcal{H}=\mathcal{H}_v\otimes\mathcal{H}_B\equiv\mathcal{H}_{v}
\otimes\mathcal{H}_B^{(a)} \otimes\mathcal{H}_B^{(b)}$, and the
state of the full system is also a direct product, $|{\bf x},{\vec
\alpha}\rangle\equiv|{\bf x}\rangle\otimes |{\vec
\alpha}\rangle\equiv|{\bf x}\rangle\otimes |{\vec
\alpha}_a\rangle\otimes|{\vec \alpha}_b\rangle$.  We use the
coordinate representation for the defect (${\bf x}$ are the
eigenvalues of its center-of-mass position operator), and the
coherent state representation for the bath, $\hat{a}_{nm}|{\vec
\alpha}_a\rangle=\alpha_{nm,a}|{\vec \alpha_a }\rangle$ and
$\hat{b}_{nm}|{\vec \alpha_b}\rangle=\alpha_{nm,b}|{\vec \alpha_b
}\rangle$.  The reduced density matrix is defined as
$\hat{\tilde{\rho}}_{v}(t)={\rm tr}_B[\hat{\rho}(t)]$, where ${\rm
tr}_B$ denotes the trace over the bath degrees of freedom, and
$\hat{\rho}(t)$ is the density matrix of the full system, whose
evolution is described by
\begin{equation}\lb{dmevol}
\hat{\rho}(t)=e^{-\frac{i\hat{H}t}{\hbar}}\hat{\rho}(0)
e^{\frac{i\hat{H}t}{\hbar}}.
\end{equation}

Here, $\hat{H}$ is the Hamiltonian of the full system given by
Eqs.\ (\ref{qham})-(\ref{Dpm}). The matrix elements of the density
operator in the basis introduced before are
$$
\hat{\rho}({\bf x},{\vec \alpha};{\bf y},{\vec \beta};t)
=\langle{\bf x},{\vec \alpha}|\hat{\rho}(t)|{\bf y},{\vec
 \beta}\rangle,
$$
and the reduced density matrix of the vortex reads
\begin{eqnarray}\lb{rdmevol1}
&&\hat{\tilde{\rho}}_{v}({\bf x},{\bf y},t)=\int
\frac{d^2{\vec \alpha}}{\pi^{2N}}\langle{\bf x},{\vec \alpha}|
\hat{\rho}(t)|{\bf y},{\vec \alpha}\rangle\nonumber\\
&=&\int\frac{d^2{\vec \alpha}}{\pi^{2N}}
\langle{\bf x},{\vec \alpha}|e^{-\frac{i\hat{H}t}{\hbar}
}\hat{\rho}(0)e^{\frac{i\hat{H}t}{\hbar}}|{\bf y},{\vec
 \alpha}\rangle.
\end{eqnarray}
After insertion of the unity operator
on both sides of $\hat{\rho}(0)$ in
Eq.\ (\ref{rdmevol1}), the reduced
density matrix acquires the form
\begin{eqnarray}&&\lb{rdmevolution}
\hat{\tilde{\rho}}_{v}=\int \frac{d^2{\vec \alpha}}{\pi^{2N}} \int
d^2{\bf x}'\int d^2{\bf y}' \int\frac{d^2{\vec
\beta}}{\pi^{2N}}\int\frac{d^2{\vec \beta}'} {\pi^{2N}}\nonumber\\
&& \langle{\bf x},{\vec \alpha}|e^{-\frac{i\hat{H}t} {\hbar}}|{\bf
x}',{\vec \beta}\rangle\langle {\bf x}',{\vec
\beta}|\hat{\rho}(0)| {\bf y}',{\vec \beta}'\rangle \langle{\bf
y}',{\vec \beta}'|e^{\frac{i\hat{H}t}
{\hbar}}|{\bf y},{\vec \alpha}\rangle.\nonumber\\
&&
\end{eqnarray}
In order to calculate the time evolution of the reduced density
matrix, we have to define the initial condition for the density
matrix of the full system. For the sake of simplicity, we choose
the factorizable one
\begin{equation}\lb{ic}
\hat{\rho}(0)=\hat{\tilde{\rho}}_{v}(0)
\hat{\tilde{\rho}}_B(0),
\end{equation}
which implies that the bath and the topological defect are
decoupled at $t=0$. The bath is assumed to be initially in
thermal equilibrium at temperature $T$,
\begin{equation}
\lb{icb}
\hat{\tilde{\rho}}_B=\frac{e^{-U\hat{H}_B}}
{{\rm tr}[e^{-U\hat{H}_B}]}=\hat{\tilde{\rho}}_{B,a}\hat{\tilde{\rho}}_{B,b},
\end{equation}
where $U\equiv\hbar/(k_BT)$. Here, we used the fact that
the baths do not interact, so the density matrix of the
full bath is the product of the density matrices
for the separate baths.  By substituting
Eqs.\ (\ref{ic}) and (\ref{icb}) into Eq.\
(\ref{rdmevolution}) we obtain
\begin{equation}
\hat{\tilde{\rho}}_{v}({\bf x},{\bf y},t)=\int d^2{\bf x}'
\int d^2{\bf y}'J({\bf x},{\bf y},t;{\bf x}',{\bf y}',0)
\hat{\tilde{\rho}}_{v}({\bf x}',
{\bf y}',0),
\end{equation}
with the superpropagator $J$ given by
\begin{eqnarray}\lb{superprop}
&&J({\bf x},{\bf y},t;{\bf x}',{\bf y}',0)=\int\frac{d^2{\vec \alpha}}{\pi^{2N}}
\int\frac{d^2{\vec \beta}}{\pi^{2N}}\int\frac{d^2{\vec \beta}'}{\pi^{2N}}
\hat{\tilde{\rho}}_B({\vec \beta}^*,{\vec \beta}',0)
\nonumber\\
&&K({\bf x},{\vec \alpha}^*;{\bf x}',{\vec \beta};t)
K^*({\bf y},{\vec \alpha}^*;{\bf y}',{\vec \beta}';t).
\end{eqnarray}
\begin{widetext}
\subsection{The Superpropagator}
We consider the superpropagator (\ref{superprop}).
The kernel
\begin{equation}\lb{defkernel}
K({\bf x},{\vec \alpha}^*;{\bf y},{\vec \beta};t)
\equiv \langle{\bf x},{\vec \alpha}|e^{-\frac{i\hat{H}t}
{\hbar}}|{\bf y},{\vec \beta}\rangle,
\end{equation}
can be expressed in the path-integral formalism as (Appendix C)
\begin{equation}\lb{kernel}
K({\bf x}, {\vec \alpha}^*;{\bf y},{\vec \beta};t)
=\int_{{\bf y}}^{{\bf x}} \mathcal{D}{\bf x}
\exp\left\{-\frac{|{\vec \alpha}|^2}{2}-\frac{|{\vec \beta}|^2}{2}\right\}
\int_{{\vec \beta}}^{{\vec \alpha}^*}\mathcal{D}
{\vec \zeta} \exp\left\{\frac{1}{2}[{\vec \zeta}^*(0)
\cdot{\vec \beta}
+{\vec \zeta}(t)\cdot{\vec \alpha}^*]\right\}
\exp\left\{\frac{i}{\hbar}S_0[{\bf x}]\right\}
\exp\left\{S_I[{\bf x},{\vec \zeta}]\right\},
\end{equation}
where $S_0[{\bf x}]=\int_0^t dt'(M/2)\dot{\bf x}^2$ stands for the
free action and
\begin{equation} \label{int5}
 S_I[{\bf x},{\vec \zeta}]=\int_0^t dt'\left\{
\frac{1}{2}({\vec \zeta}\cdot\dot{{\vec \zeta}^*}-
{\vec \zeta}^*\cdot\dot{{\vec \zeta}})-
\frac{i}{\hbar}\left[h_B({\vec \zeta}^*,
{\vec \zeta})-
\dot{{\bf x}}\cdot{\bf h}_I({\vec \zeta}^*,
{\vec \zeta})\right]\right\}\equiv S_{I,a}[{\bf x},{\vec \zeta_a}]+S_{I,b}[{\bf x},{\vec \zeta_b}]
\end{equation}
denotes the interaction. Here
$$
h_B=\sum_{nm,i=a,b}\hbar\omega_{nm}\zeta_{nm,i}^*\zeta_{nm,i}\equiv h_{B,a}+h_{B,b},\qquad
{\bf h}_I=\hbar\sum_{nm,kl}{\bf D}_{nm,kl}\zeta_{kl,a}^*
\zeta_{nm,a}-{\bf D}_{kl,nm}\zeta_{nm,b}\zeta_{kl,b}^*
\equiv {\bf h}_{I,a}+{\bf h}_{I,b}.
$$
By inserting Eq.\ (\ref{kernel}) into Eq.\ (\ref{superprop}) and
using the reduced density matrix of the bath in the coherent state
representation (Appendix D), we obtain a simpler expression for
the superpropagator,
\begin{equation}\lb{superprop1}
J({\bf x},{\bf y},t;{\bf x}',{\bf y}',0)=
\int_{{\bf x}'}^{\bf x}\mathcal{D}{\bf x}
\int_{{\bf y}'}^{\bf y}\mathcal{D}{\bf y}e^{\frac{i}
{\hbar}[S_0({\bf x})-S_0({\bf y})]} \mathcal{F}[{\bf x},{\bf y}],
\end{equation}
where $\mathcal{F}=\mathcal{F}_a\mathcal{F}_b$ is the total influence functional,
and $\mathcal{F}_i$ ($i=a,b$) is the influence functional  for the
bath $i$   given by (to simplify notation we omit the index $i$ in the
integration variables)
\begin{eqnarray}\lb{influence5}
&&\mathcal{F}_i[{\bf x},{\bf y}]=
\int\frac{d^2{\vec \alpha}}{\pi^N}
\int\frac{d^2{\vec \beta}}{\pi^N}
\int\frac{d^2{\vec \beta}'}{\pi^N} \rho_B({\vec \beta}^*
,{\vec \beta}',0) \exp\left\{-|{\vec \alpha}|^2-
\frac{|{\vec \beta}|^2}{2}-\frac{|{\vec \beta}'|^2}{2}
\right\}\nonumber\\
&&\int_{\vec \beta}^{{\vec \alpha}^*}
\mathcal{D}{\vec \zeta}\int_{{\vec \beta}'^*}^{{\vec \alpha}}
\mathcal{D}{\vec \gamma}
\exp\left\{\frac{1}{2}\left[{\vec \zeta}^*(0)\cdot
{\vec \beta}+{\vec \zeta}(t)\cdot
{\vec \alpha}^*
+{\vec \gamma}(0)\cdot{\vec \beta}'^*
+{\vec \gamma}^*(t)\cdot{\vec \alpha}\right]\right\}
\exp\left\{S_{I,i}[{\bf x},{\vec \zeta}]+
S_{I,i}^*[{\bf y},{\vec \gamma}]\right\},
\end{eqnarray}
with the initial conditions
\begin{equation}\lb{ic1}
{\vec \zeta}(0)={\vec \beta},\qquad {\vec \zeta}^*(t)=
{\vec \alpha}^*,
\end{equation}
\begin{equation}\lb{ic2}
{\vec \gamma}^*(0)={\vec \beta}'^*\qquad
{\vec \gamma}(t)={\vec \alpha}.
\end{equation}
\subsection{Influence Functional}
We now evaluate the influence functional, which describes the
influence of the bath on the effective dynamics of the defect. The
only difference between the functionals $\mathcal{F}_a$ and
$\mathcal{F}_b$ is in the form of the interaction $S_I$; see Eqs.
(\ref{int5}) and (\ref{influence5}). Note that the actions
$S_{I,a}$ and $S_{I,b}$ are related by the substitution ${\bf
D}_{nm,kl} \rightarrow -{\bf D}_{kl,nm}$. Thus, it is enough to
calculate the functional $\mathcal{F}_a$, and consequently
$\mathcal{F}_b$ is obtained using the latter transformation. In
order to simplify notation, in what follows we write the
integration variables without the index $a$. First, we calculate
the path-integrals in Eq.\ (\ref{influence5}) using the stationary
phase approximation (SPA). In order to apply the SPA, we have to
solve the equations of motion corresponding to $S_I$ and $S_I^*$.
Because $S_I^*({\bf x},{\vec \zeta})= -S_I({\bf x},{\vec \zeta})$,
we need to consider only $S_I$. The equations of motion are
promptly
 obtained from
$\delta S_I/\delta\zeta^*_{nm}=0$,
$\delta S_I/\delta\zeta_{nm}=0$, and they read
\begin{eqnarray}\lb{eqm1}
&&\dot{\zeta}_{nm}+i\omega_{nm}\zeta_{nm}-i\dot{\bf x}
\sum_{kl}{\bf D}_{kl,nm}\zeta_{kl}=0,
\nonumber\\
&&\dot{\zeta}^*_{nm}-i\omega_{nm}\zeta^*_{nm}+i\dot{\bf x}
\sum_{kl}{\bf D}_{nm,kl}\zeta^*_{kl}=0.\nonumber\\
\end{eqnarray}
Notice that the two equations are identical, one is the
complex conjugate of the other (recall that
${\bf D}_{nm,kl}={\bf D}^*_{kl,nm}$).

The SPA requires the evaluation of the action $S_I$ on the
classical trajectory, which is the solution of the above equations
of motion. Straightforward
calculations show that the value of $S_I$ at the stationary point
is zero. If we define ${\vec \zeta}={\vec \zeta}_{cl}
+{\vec \Delta}$, then the functional integral
over ${\vec \zeta}$ becomes the functional
integral over the fluctuations ${\vec \Delta}$ around the
saddle point. Expanding the action around its saddle point,
we find that the relevant contribution
comes from the second
derivative of the action at the stationary point, because
both the value and the first derivative of
the action are zero at the saddle point. The second
derivative of the action evaluated at the
stationary point is a constant operator,
so the integration over the fluctuations $\Delta$ yields,
up to an irrelevant constant,
\begin{widetext}
\begin{equation}\lb{influence1}
\mathcal{F}_a[{\bf x},{\bf y}]=\int\frac{d^2{\vec \alpha}}{\pi^N}
\int\frac{d^2{\vec \beta}}{\pi^N}\int\frac{d^2{\vec \beta}'}{\pi^N}
\rho_{B,a}({\vec \beta}^*,{\vec \beta}')\rm{e}^{-|{\vec \alpha}|^2-
\frac{|{\vec \beta}|^2}{2}-\frac{|{\vec \beta}'|^2}{2}}
\exp\left\{\frac{1}{2}\left[{\vec \zeta}^*(0)\cdot
{\vec \beta}+{\vec \zeta}(t)\cdot
{\vec \alpha}^*+{\vec \gamma}(0)\cdot{\vec \beta}'^*
+{\vec \gamma}^*(t)\cdot{\vec \alpha}\right]\right\}.
\end{equation}
\end{widetext}
Therefore, in the SPA, $S_I$ and $S_I^*$  only contribute to the
influence functional through the boundary terms, which may be
determined using the solutions of the equations of motion (see
Appendix E).

After inserting Eqs.\
(\ref{cohbdm}) and (\ref{gammat}) into Eq.\ (\ref{influence1}), and performing
the Gaussian integrals over $\alpha,\beta$ and $\beta'$,
the influence functional acquires the form
\begin{equation}\lb{influence}
\mathcal{F}_a[{\bf x},{\bf y}]=\frac{1}{{\rm det}(1-\bar{n}\Gamma^a)},
\end{equation}
where the matrix $\Gamma^a$ is given by
\begin{eqnarray}\lb{Gamma}
&&\Gamma_{nm,kl}^a[{\bf x},{\bf y}]=\frac{1}{2}[W_{kl,nm}({\bf x},t)+
\tilde{W}_{nm,kl}({\bf x},0)
\nonumber \\
&+&  \bar{W}_{kl,nm}({\bf y},0)
+\tilde{\bar{W}}_{nm,kl}({\bf y},t)]
+\frac{1}{4}\sum_{pq}[\tilde{W}_{nm,kl}({\bf x},0)
\nonumber \\
&+&
W_{pq,nm}({\bf x},t)]
[\bar{W}_{kl,pq}({\bf y},0)+
\tilde{\bar{W}}_{pq,kl}({\bf y},t)],
\end{eqnarray}
and $\bar{n}_{pq}=1/[\exp(U\omega_{pq})-1]$ is the bosonic
occupation number. Equation\ (\ref{tildewwstar}) enables us to
express the matrix $\Gamma^a$ only in terms of the functionals $W$
and $\tilde{W}$
\begin{eqnarray}\lb{Gamma1}
&&\Gamma^a_{nm,kl}[{\bf x},{\bf y}]=\frac{1}{2}[W_{kl,nm}
({\bf x},t)+\tilde{W}_{nm,kl}({\bf x},0)
\nonumber \\
&+&
\tilde{W}_{kl,nm}^*({\bf y},0)+W^*_{nm,kl}({\bf y},t)]
+\frac{1}{4}\sum_{pq}[\tilde{W}_{nm,pq}({\bf x},0)
\nonumber \\
&+& W_{pq,nm}({\bf x},t)][\tilde{W}^*_{kl,pq}({\bf y},0)
+ W^*_{pq,kl}({\bf y},t)].
\end{eqnarray}
Using the formula $\ln {\rm det} \mathcal{A}={\rm tr}\ln
\mathcal{A}$ for the matrix
$\mathcal{A}=(1-\bar{n}\Gamma^a)^{-1}$, we find
\begin{equation}\lb{ngama}
\mathcal{F}_a[{\bf x},{\bf y}]=\exp[{\rm tr} (\bar{n}\Gamma^a)]=
\exp\left[\sum_{pq}\bar{n}_{pq}\Gamma^a_{pq,pq}\right],
\end{equation}
and the total influence functional reads
\begin{equation}
\mathcal{F}=\mathcal{F}_a\mathcal{F}_b=
\exp\left[\sum_{pq}\bar{n}_{pq}\Gamma_{pq,pq}\right]
\end{equation}
in the lowest order in $\bar{n}\Gamma$, where $\Gamma\equiv
\Gamma^a+\Gamma^b$. The diagonal elements of the matrix $\Gamma^a$
are obtained from Eq.\ (\ref{Gamma1}), while the matrix $\Gamma^b$
is obtained from  $\Gamma^a$ by the substitution ${\bf
D}_{nm,kl}\rightarrow -{\bf D}_{kl,nm}=-{\bf D}^*_{nm,kl}$. The
functionals $W$ and $\tilde{W}$ are given implicitly by Eqs.\
(\ref{W1}). From their form we see that they actually represent
the amplitude of scattering of the mode $nm$ to the mode $kl$
through virtual intermediate states. These functionals can be
determined iteratively from these equations up to any order. Here,
we study the motion of a vortex with small kinetic energy;
therefore the Born approximation will be enough for our purpose.
The functionals $W$ and $\tilde{W}$ are calculated within the Born
approximation in Appendix E. Using Eq.\ (\ref{BtW}) the diagonal
elements of the matrix $\Gamma$ can be promptly evaluated
(Appendix F), and the total influence functional reads
\begin{equation}\lb{influence2}
\mathcal{F}[{\bf x},{\bf y}]=\exp[\frac{i}{\hbar}\Phi]\exp[\tilde{\Phi}],
\end{equation}
where
\begin{widetext}
\begin{eqnarray}\lb{Phi}
&&\Phi=\sum_{\mu,\nu=1}^2\int_{0}^{t}dt'\int_{0}^{t}dt''\theta(t'-t'')
\epsilon^{\mu\nu}(t'-t'') [\dot{x}^{\mu}(t')-\dot{y}^{\mu}(t')][\dot{x}^{\nu}(t'')+\dot{y}^{\nu}(t'')],
\end{eqnarray}
$$
\tilde{\Phi}=\sum_{\mu\nu}\int_{0}^{t}dt'\int_{0}^{t}dt''\theta(t'-t'')
\tilde{\epsilon}^{\mu\nu}(t'-t'')[\dot{x}^{\mu}(t')-\dot{y}^{\mu}(t')][\dot{x}^{\nu}(t'')-\dot{y}^{\nu}(t'')]
$$
with
\begin{eqnarray}\lb{epsilon}
&&\epsilon^{\mu\nu}(t)=-\hbar\sum_{\mu,\nu}\sum_{nm,kl}\bar{n}_{nm} \left(D^{\mu*}_{nm,kl} D^{\nu}_{nm,kl}+
D^{\mu}_{nm,kl} D^{\nu*}_{nm,kl}\right)
\sin(\omega_{nm}-\omega_{kl})t,
\end{eqnarray}
\begin{eqnarray}
&&\tilde{\epsilon}^{\mu\nu}(t)=-\sum_{\mu,\nu}\sum_{nm,kl}\bar{n}_{nm} \left(D^{\mu*}_{nm,kl} D^{\nu}_{nm,kl}
+ D^{\mu}_{nm,kl} D^{\nu*}_{nm,kl}\right)
\cos(\omega_{nm}-\omega_{kl})t.
\nonumber
\end{eqnarray}
\end{widetext}
From Eqs.\ (\ref{superprop1}) and (\ref{influence2}) we see that
the oscillatory part $\exp[i\Phi/\hbar]$ gives a contribution to
the effective action of the defect  due to its scattering by the
magnons  and leads to its dissipative motion, as we show in the
following section. The decaying part $\exp[\tilde{\Phi}]$ is
related to the diffusive properties of the vortex. The diffusive
and damping properties of the defect  are related at low
temperatures by the fluctuation-dissipation theorem.
\end{widetext}

\section{The dynamics of the defect}
\subsection{Transport properties of the defect}
In this section we shall study the effective dynamics of the
defect after integrating out the magnons. According to  Eqs.\
(\ref{superprop1}) and (\ref{influence2}) the effective action
describing the influence of magnons on the motion of the
topological defect reads
\begin{equation}
\lb{effaction}
S_{eff}=S_0[{\bf x}]-S_0[{\bf y}]+\Phi[{\bf x},{\bf y}],
\end{equation}
where $\Phi$ is given by Eq.\ (\ref{Phi}). Since $\Phi \propto
{\bf D}$ we observe that if the coupling constants ${\bf D}$ were
zero, then the motion of the defect would be free. The equations
of motion for the defect can be directly obtained by extremizing
the effective action (\ref{effaction}), $\delta S_{eff}/\delta
x^\mu=0$ and $\delta S_{eff}/\delta y^\mu=0$. In terms of the
center of mass ${\bf v}=({\bf x}+{\bf y})/2$ and relative ${\bf
u}={\bf x}-{\bf y}$ coordinates, they read
\begin{eqnarray}\lb{eqsmotion}
&&\ddot{v}^\mu(\tau)+\sum_\nu\int_0^\tau dt'
\gamma^{\mu\nu}(\tau-t')\dot{v}^\nu(t')=0,\nonumber\\
&&\ddot{u}^\mu(\tau)+\sum_\nu\int_\tau^t dt'
\gamma^{\mu\nu}(t'-\tau)\dot{u}^\nu(t')=0.
\end{eqnarray}
As a consequence, the damping
matrix $\gamma^{\mu\nu}$ is given by
\begin{widetext}
\begin{eqnarray}
&&\gamma^{\mu\nu}(t)\equiv\frac{2}{M}\frac{d}{dt}
\epsilon^{\mu\nu}(t)=-\frac{\hbar}{M}\sum_{nm,pl}
(\bar{n}_{nm}-\bar{n}_{pl})(\omega_{nm}-\omega_{pl})
\left(D^{\mu*}_{nm,pl}D^{\nu}_{nm,pl}+
D^{\mu}_{nm,pl}D^{\nu*}_{nm,pl}\right)\cos(\omega_{nm}-\omega_{pl})t\nonumber\\
\lb{damping} &=&-\frac{\hbar}{4M} \sum_{nm,pl}(\bar{n}_{nm}-\bar{n}_{pl})
\frac{(\omega_{nm}-\omega_{pl})(\omega_{nm}+\omega_{pl})^2}
{\omega_{nm}\omega_{pl}}
\left(G^{\mu*}_{nm,pl}G^\nu_{nm,pl}+
G^{\mu}_{nm,pl}G^{\nu*}_{nm,pl}\right) \cos(\omega_{nm}-\omega_{pl})t\nonumber\\
&=&-\frac{\hbar}{2M} \sum_{nm,pl}(\bar{n}_{nm}-\bar{n}_{pl})
\frac{(\omega_{nm}-\omega_{pl})(\omega_{nm}+\omega_{pl})^2}
{\omega_{nm}\omega_{pl}}G^{\mu*}_{nm,pl}G^\nu_{nm,pl}
\cos(\omega_{nm}-\omega_{pl})t,
\end{eqnarray}
where $\epsilon^{\mu\nu}$ was defined in Eq.\ (\ref{epsilon}).
Introducing the scattering matrix
\begin{equation}
\lb{smatrix}
S^{\mu\nu}(\omega,\omega')=\sum_{nm,pl} G^{\mu*}_{nm,pl}
G^\nu_{nm,pl}\delta(\omega-\omega_{nm})\delta(\omega'- \omega_{pl})
\end{equation}
and observing that because of the isotropy of the model  the
damping matrix is diagonal (see also Eq.\ (\ref{Smatrix}) below),
$\gamma^{\mu\nu}(t)=\gamma(t)\delta^{\mu\nu}$, the damping
function $\gamma(t)$ is given by
\begin{equation} \lb{defdampmatrix}
\gamma(t)=-\frac{\hbar}{2M}\int_{0}^\infty
d\omega\int_{0}^\infty d\omega'[\bar{n}(\omega)
-\bar{n}(\omega')]
\frac{(\omega-\omega')
(\omega+\omega')^2}{\omega\omega'}S
(\omega,
\omega')\cos(\omega-\omega')t.
\end{equation}
\end{widetext}
Let us introduce the new variables $\xi=(\omega+\omega')/2$,
$\psi=\omega-\omega'$. The damping function can be rewritten as
\begin{equation}
\lb{dampingcoefficient}
\gamma(t)=\int_{-\infty}^\infty d\psi \mathcal{J}(\psi)\cos\psi t,
\end{equation}
where $\mathcal{J}(\psi)$ is the spectral function of the bath
\cite{amir}  given by an additional integration
\begin{eqnarray}
\mathcal{J}(\psi)&=&-\frac{2\hbar}{M}\int_0^\infty
d\xi\left[\bar{n}\left(\xi+\frac{\psi}{2}\right)
-\bar{n}\left(\xi-\frac{\psi}{2}\right)\right]
\nonumber\\
\lb{rpsi} &&\times\frac{\xi^2\psi}{\xi^2-\psi^2/4}S\left(\xi+
\frac{\psi}{2},\xi-\frac{\psi}{2}\right).
\end{eqnarray}
From the equations of motion (\ref{eqsmotion}) it is easy to see
that if a charge $q$ is associated with the defect (see next
section) the corresponding optical conductivity is
\begin{equation}
\lb{optical}
\sigma(\omega)=\left .\frac{nq^2/M}{z+\hat\gamma(z)}\right|_{z\rightarrow
-i\omega+0^+},
\end{equation}
where $n$ is the density of carriers and $\hat\gamma(z)$ is the Laplace
transformation of the damping function
\begin{equation}
\lb{lap}
\hat\gamma(z)=\int_0^\infty dt \gamma(t)e^{-zt}=
\int_{-\infty}^\infty d\psi\frac{z \mathcal{J}(\psi)}{z^2+\psi^2}.
\end{equation}
When only quasielastic processes are taken into account
$\psi\approx 0$, so that we can approximate
$\mathcal{J}(\psi)\approx \mathcal{J}(0)$, and using the fact the
$\int_{-\infty}^{\infty} d\psi\cos\psi t=2\pi\delta(t)$, we find
from Eqs.\ (\ref{dampingcoefficient}) and (\ref{rpsi}) that
$$
\gamma(t)=\bar\gamma\delta(t), \quad  \hat\gamma(z)=\bar\gamma/2,
$$
with the damping coefficient $\bar\gamma(T)$
\begin{equation}
\lb{barg}
\bar\gamma(T)=2\pi \mathcal{J}(0)=-\frac{4\pi\hbar}{M}\int_0^\infty
d\xi\varphi(\xi)\frac{\partial\bar{n}(\xi)}{\partial\xi},
\end{equation}
where we defined
\begin{equation}\lb{limit}
\varphi(\xi)=\lim_{\psi\to0}\psi^2
S\left(\xi+\frac{\psi}{2},\xi-\frac{\psi}{2}\right).
\end{equation}
According to Eq.\ (\ref{optical}) the real part of the optical conductivity
has a typical Drude-like shape
\begin{equation}
\lb{drude}
Re\sigma(\omega)=\frac{nq^2}{M}\frac{\bar\gamma/2}{\omega^2+(\bar\gamma/2)^2},
\end{equation}
where $\bar\gamma/2$ plays the role of the inverse scattering time. It is
then natural to introduce the inverse mobility as
\begin{equation}
\lb{defmu}
\mu^{-1}=\frac{M\bar\gamma}{2q}.
\end{equation}
It is worth noting that even though the formula (\ref{optical}) is
general, in the computation of the damping function
(\ref{damping}), we considered only low-energy  quasielastic
processes, which, naturally, should lead to the Drude-like
response (\ref{drude}). If one keeps in the evaluation of
$\mathcal{J}(\psi)$ the next nonvanishing contribution to Eq.\
(\ref{lap}),
$$
\mathcal{J}(\psi)\approx \mathcal{J}(0)+\mathcal{J}_2\psi^2, \quad \psi<\omega_c,
$$
where $\omega_c$ is a proper cutoff for the previous expansion,
$\hat\gamma(z)$ can be estimated as
$$
\hat\gamma(z)=\pi\left[\mathcal{J}(0)-z^2 \mathcal{J}_2\right]+
2\mathcal{J}_2\omega_c z.
$$
The corresponding optical conductivity is then
$$
Re\sigma(\omega)=\frac{nq^2}{M}\frac{\pi \mathcal{J}(0)+\pi \mathcal{J}_2\omega^2}
{\omega^2(1+2\mathcal{J}_2\omega_c)^2+(\pi \mathcal{J}(0)+\pi \mathcal{J}_2\omega^2)^2},
$$
which is qualitatively the same as the one given in Eq.\
(\ref{drude}). In particular, the dc conductivity is found to be
the same. Since in the following we will address the issue of the
temperature dependence of the resistivity, we can safely rely on
the approximation (\ref{barg}) of the damping coefficient, which
takes into account only the contribution of $\mathcal{J}(0)$.

\subsection{Evaluation of the damping coefficient}

In order to determine  the damping coefficient (\ref{barg}), we
first have to evaluate the function $\varphi(\xi)$ defined in Eq.\
(\ref{limit}). By rewriting the summations over the (radial)
indexes $n,p$ in Eq.\ (\ref{smatrix}) as integration over
continuum variables, and taking into account that $\Delta q_n\sim
\pi/\ell$ we find that
\begin{eqnarray}\lb{Smatrix} &&
S(k,k')\equiv S^{yy}(k,k')=S^{xx}(k,k')\nonumber\\
&=&\frac{\ell^2}{\pi^2c^2}\sum_{ml}\int dq\int dq'
|G^{x}_{qm,q'l}|^2\delta(k-q)
\delta(k'-q')\nonumber\\
&=&\frac{\ell^2}{\pi^2c^2}\sum_{ml}
|G^{x}_{km,k'l}|^2,\nonumber\\
&&S^{xy}(k,k')=\frac{\ell^2}{\pi^2c^2}\sum_{ml}G^{x*}_{km,k'l}G^{y}_{km,k'l}=0,
\end{eqnarray}
where $k=\omega/c$, $k'=\omega'/c$ and the last equation follows
from Eq.\ (\ref{explcc}). The limit (\ref{limit}) thus acquires
the form
\begin{equation}\lb{limit1}
\varphi(\xi)=\frac{\ell^2}{\pi^2}\lim_{(k-k')\to0}(k-k')^2\sum_{ml}
|G_{mk,lk'}^x|^2.
\end{equation}
From the above relation it is obvious that the only terms of
$|G_{km,k'l}^x|^2$ which contribute to the limit (\ref{limit}) are
those behaving like $\sim 1/(k-k')^2$, i.e., the term proportional
to $\Lambda^{(2)}$ in Eq.\ (\ref{explcc}). We then find
\begin{equation}\lb{G2}
\sum_{ml}|G_{km,k'l}^{(2)}|^2=\frac{2k^2}{\ell^2 (k-k')^2}\sum_m
|\Lambda_{m+1,m}^{(2)}|^2,
\end{equation}
where
\begin{eqnarray}\lb{lambda}
&&\left|\Lambda_{m+1,m}^{(2)}\right|^2=
\left(e^{i\pi/2}+e^{-i\pi/2}e^{2i(\delta_{m+1}-
\delta_m)}\right)\left(e^{-i\pi/2}\right.\nonumber\\
&+&\left. e^{i\pi/2}e^{-2i(\delta_{m+1}-
\delta_m)}\right)=4\sin^2(\delta_{m+1}-\delta_m).
\end{eqnarray}
Substituting Eq.\ (\ref{lambda}) into Eq.\ (\ref{G2}), and evaluating the
limit (\ref{limit1}) we finally obtain
\begin{equation}
\varphi(\xi)=\frac{8\xi^2}{\pi^2c^2}\mathcal{G}(\xi),
\end{equation}
where $\xi=ck$ and we defined
\begin{equation}
\lb{defg}
\mathcal{G}\equiv\sum_{m=0}^\infty\sin^2(\delta_{m+1}-
\delta_m).
\end{equation}
Using the preceding equation, the damping coefficient (\ref{barg})
acquires the form
\begin{equation}
\lb{barg2}
\bar{\gamma}(T)=\frac{32\hbar^2}{M\pi c^2k_BT}\int_0^\infty
d\xi\mathcal{G}(\xi)\frac{\xi^2e^{\frac{\hbar\xi}{k_BT}}}
{\left(e^{\frac{\hbar\xi}{k_BT}}-1\right)^2}.
\end{equation}
Observe that Eq.\ (\ref{barg2}) is valid for both kinds of defect solution
(\ref{1v}) and (\ref{2v}). However, since the phase shifts are determined
by the eigenfunction $\eta$ of the scattering problem (\ref{schroedinger}),
$\Psi_{1v}$ or $\Psi_{2v}$ will give rise to  different potentials $V_{1v}({\bf r})$ or $V_{2v}({\bf r})$, and
then different phase shifts. As a consequence, also the function ${\cal
G}(\xi)$ in Eq.\ (\ref{defg}) will be different in the two cases, leading
to a different temperature dependence of the damping coefficient
(\ref{barg2}).

\section{Inverse mobility}
We evaluate the phase shifts by adopting the Born approximation.
\cite{morse,gottfried}  The phase shift $\delta_m(k)$ of the
wavefunction with angular momentum $m$ and wave vector $k$ then
reads
\begin{equation}
\lb{defd}
\delta_m(k)=\arctan \pi \mathcal{A}_m(k),
\end{equation}
where $\mathcal{A}_m(k)$ is the expectation value of the potential
over the eigenfunction of the corresponding unperturbed Schr\"
odinger equation, i.e., the Bessel function $J_m(kr)$ in the case
of Eq.\ (\ref{schroedinger})
\begin{equation}\lb{defAm}
\lb{defa}
\mathcal{A}_m(k)= \int_0^\infty
drr [J_m(kr)]^2V(r).
\end{equation}

\subsection{Vortex-antivortex pair}
Let us first consider the case of the scattering of a
vortex-antivortex pair by the magnons. In this case the potential
$V({\bf r})$ in Eq.\ (\ref{schroedinger}) has the form
\begin{equation}
\lb{v2v}
V_{2v}({\bf r})=\frac{1}{4}(\nabla\Psi_{2v})^2,
\end{equation}
where $\Psi_{2v}=\Psi_{1v}^{(1)}-\Psi_{1v}^{(2)}$, with $\Psi_{1v}$ given
by Eq.\ (\ref{sv}). One can readily show that
\begin{eqnarray}
\left(\nabla\Psi_{1v}^{(1,2)}\right)^2&=&\frac{1}
{({\bf r}-{\bf R}_{1,2})^2},\nonumber\\
\left(\nabla\Psi_{1v}^{(1)}\right)
\left(\nabla\Psi_{1v}^{(2)}\right)&=&\frac{({\bf r}-
{\bf R}_1)\cdot({\bf r}-{\bf R}_2)}
{({\bf r}-{\bf R}_1)^2({\bf r}-{\bf R}_2)^2},\nonumber
\end{eqnarray}
which gives, using the translational invariance ${\bf r}-{\bf
R}\rightarrow {\bf r}$
$$
(\nabla\Psi_{2v})^2=\frac{{\bf d}^2}
{({\bf r}-{\bf d}/2)^2({\bf r}+{\bf d}/2)^2}=
\frac{d^2}{(r^2+d^2/4)^2-({\bf r}\cdot{\bf d})^2}.
$$
Since the distance $d$ between defects is a fixed parameter and we
consider length scales $r\gg d$, we can approximate the potential
as
\begin{equation}\lb{potential2v}
V_{2v}({\bf r})=\frac{d^2/4}{(r^2+d^2/4)^2}.
\end{equation}
For the potential (\ref{potential2v}), Eq.\ (\ref{defa}) can be
evaluated  analytically for $m\geq 1$, yielding
\begin{eqnarray}
\mathcal{A}_m(k)&=&\frac{\pi kd}{4}\left\{I_m(kd/2)
\left[K_{m-1}(kd/2)
+K_{m+1}(kd/2)\right]\right.\nonumber\\
&-&\left. K_m(kd/2)[I_{m-1}(kd/2)
+I_{m+1}(kd/2)]\right\},\nonumber
\end{eqnarray}
where $I_n$ and $K_n$ are the modified Bessel functions of the
first and the second kinds, respectively.

Returning to Eq.\ (\ref{defg}), we observe that ${\cal
G}(\xi=k/c)$ is a function of the dimensionless variable $y=\xi
d/2c$. Introducing the same variable also into Eq.\ (\ref{barg2}),
we may rewrite the inverse mobility given by Eq.\ (\ref{defmu}) as
\begin{equation}
\lb{mu} \mu^{-1}(T)=\mu^{-1}_0
\frac{128}{\pi}\frac{E_c}{\alpha^3T} \int_0 ^\infty dy
y^2\mathcal{G}(y) \frac{e^{\frac{2E_cy}{\alpha T}}}
{(e^{\frac{2E_cy}{\alpha T}}-1)^2},
\end{equation}
where $\mu_0^{-1}=\hbar/ e a^2$ is the quantum of inverse mobility
for a given lattice spacing $a$, $d=\alpha a$ and $E_c=\hbar c/a
k_B$ is the characteristic temperature scale associated with the
magnons. Even though a quantitative estimate of Eq.\ (\ref{mu})
requires the knowledge of the values of these microscopic
parameters, its qualitative behavior can be promptly understood.
In particular, since all the temperature dependence of $\mu^{-1}$
is due to the Bose factor in Eq.\ (\ref{mu}), one can expect that
the inverse mobility vanishes at zero temperature, where no
thermally activated scattering processes exist, and increases
linearly  at high temperatures, with the slope determined by the
shape of the function ${\cal G}(y)$. In the left panel of Fig.\ 5
we plot $(\mu/\mu_0)^{-1}$ as a function of $T/E_c$ for several
values of $\alpha$. One observes that already at small fractions
of the ratio $T/E_c$, the inverse mobility is {\em linear} in
temperature. Moreover,  as $\alpha$ increases the linear behavior
arises at even smaller temperatures, and the overall value of the
inverse mobility decreases.

\begin{figure}[htb]
\begin{center}
\includegraphics[width=6cm,angle=-90]{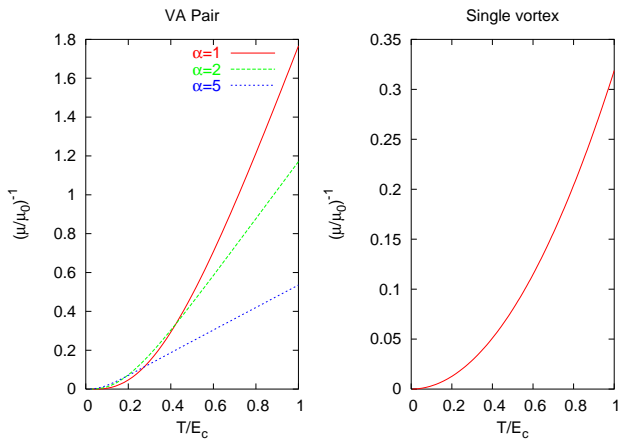}
\end{center}
\caption{(Color online) Inverse mobility in units of $\mu_0^{-1}$
as a function of the rescaled temperature $T/E_c$. Left panel:
inverse mobility of a vortex-antivortex pair, according to Eq.\
(\ref{mu}), at several values of $\alpha$. Right panel: inverse
mobility of a single-vortex defect, according to Eq.\
(\ref{musd})}
\end{figure}

\subsection{Single vortex}

Let us analyze now the behavior of the inverse mobility obtained
when we identify the defect as a single vortex. In this case,
instead of Eq.\ (\ref{v2v})  the scattering potential reads
$$
V_{1v}({\bf r})=\frac{1}{4}(\nabla \Psi_{1v})^2=\frac{1}{4r^2}.
$$
As a consequence, the phase shifts defined by Eqs.\
(\ref{defd}) and (\ref{defa}) are given by ($m\geq 1$)
\begin{equation}
\delta_m=\arctan\frac{\pi}{8m}.
\end{equation}
Note that the phase shifts in the case of a single-vortex do not
depend on the wave-vector, but only on the angular momentum. Thus,
the function $\mathcal{G}$ defined by Eq.\ (\ref{defg}), does not
depend on the frequency
$$
\mathcal{G}=\sum_{m=1}^\infty\frac{64\pi^2}{64\pi^2+
[\pi^2+64m(m+1)]^2}\approx 0.032.
$$
Because $\cal G$ is a constant, we can introduce the rescaled
variable $y\equiv\hbar \xi/k_B T$ into the expression
(\ref{defmu}) for the inverse mobility, which yields
\begin{equation}
\lb{musd} \mu^{-1}(T)=\mu_0^{-1}\frac{8\pi}{3}{\cal G}
\left(\frac{T}{E_c}\right)^2.
\end{equation}
Here, we used the fact that $\int_0^\infty dy
y^2/(e^y-1)^2=\pi^2/3$. In comparison with the case of a
vortex-antivortex pair, the main difference is that here the
inverse mobility depends on the square of the temperature, for all
the temperatures. In the right panel of Fig. 5 we plot
$(\mu/\mu_0)^{-1}$ as a function of $T/E_c$: notice that the
overall variation of the inverse mobility is smaller compared to
the case of the vortex-antivortex pair, but they are still of the
same order of magnitude.

\section{Application of the model: the case of cuprates}

\subsection{The spiral state in cuprates}

Although the model we have developed above could be applied to
describe the dynamics of topological defects in several frustrated
Heisenberg spin systems, here we concentrate on lightly doped
cuprates. Indeed, a large part of the literature devoted to
frustrated spin systems is connected to the $t-J$ model, which is
the strong-coupling limit of the Hubbard model. The latter is
considered to be the prototype of an effective description of the
CuO$_2$ planes of cuprate superconductors.  At half-filling the
$t-J$ model describes a spin-$1/2$ antiferromagnet which is
believed to have long-range order at zero temperature. As the
system is doped away from half-filling the motion of a hole will
leave a trail of spins pointing in the wrong direction. Thus, two
issues must be settled: (i) the character of the quasiparticle
wave function and (ii) the effect of the hole motion on the spin
background. Both issues have been extensively addressed in the
literature. In the atomic limit ($t=J=0$) of the Hubbard model,
considered by Brinkman and Rice, \cite{brinkman} the ``string'' of
perturbed spins can be healed only by retracing the original path
and returning all the spins to their original position. As a
consequence, in the presence of a finite but small $J$, they
argued that the ground state of the hole involves a magnetic
polaron: the cost of creating a ferromagnetic region around the
hole is compensate by the fact that inside this ferromagnetic
cloud it can sit at the free-particle band edge.

At larger value of $J$ ($1\gtrsim J/t \gtrsim 5 \times 10^{-3}$)
Shraiman and Siggia \cite{siggia881} showed that at least in the
Ising limit ($J_z\neq 0, J_\perp=0$) the picture of band narrowing
effect is more appropriate than the polaron formation. In the
Ising limit the holes are infinitely massive, because they are
self-trapped to their original position by the ``string'' of
overturned spins.  When a finite $J_\perp$ is included, quantum
spin fluctuations associated with it can repair a pair of
overturned spins and the mass of the holes becomes large but
finite.\cite{trugman} Several calculations
\cite{schmitt,kane,fulde} were performed using an effective
Hamiltonian which couples the holes (constrained to no double
occupancy) to Holstein-Primakoff spin-waves. The result is that
the hole moves on a given sublattice, forming a narrow ($\sim J$)
quasiparticle band with the minimum at the wave-vector
$(\pm\pi/2,\pm\pi/2)$, plus an incoherent part originating from
the spin-wave excitations created by the hole motion.  By using a
semiclassical approach, Shraiman and Siggia
\cite{siggia881,siggia882,siggia89} showed that one can assign to
the hole states a dipolar momentum ${\bf p}_a$, which is a vector
both in lattice and spin space. The coupling between this dipolar
moment and the magnetization current ${\bf j}_a={\bf \Omega}\times
\partial_a{\bf \Omega}$ of the antiferromagnetic background,
described by a NL$\sigma$ model for ${\bf \Omega}$, leads, at
finite doping, to a spiral re-ordering of the antiferromagnetic
phase of the background spins. Within a similar approach, Gooding
\cite{gooding} has argued that a strong localization of the hole
could eventually lead to a skyrmion-like configuration of the
background spins. However, while the polaron or the skyrmion
formation seem to be plausible scenarios for a single defect, at
finite doping the picture proposed by Shraiman and Siggia of a new
helical spin configuration is more likely. Later, many
calculations on the $t-J$ (or $t-t'-J$) model based on different
approaches have indeed confirmed that a spiral ground state can be
favorable at low doping. \cite{kane90,schulz,mori,weng,kotov}

Recently, the interest on  spiral formation in $t-J$-based models
has been revived due to the experimental observation of {\em
incommensurate} spin correlations in cuprates, i.e., an
enhancement of the spin susceptibility at a wave vector ${\bf
k}_s$ slightly displaced with respect to the commensurate wave
vector $(\pi/a,\pi/a)$. In particular, detailed measurements of
the incommensurability as a function of doping are available for
lanthanum-based compounds. \cite{NSSG,antonio} For doping $x$
larger than $0.05$, i.e., in the regime where the samples are
superconducting, the observation of four peaks at $(\pi/a\pm
\delta,\pi/a)$ and $(\pi/a,\pi/a\pm\delta)$, and the simultaneous
measurement of incommensurate charge peaks, lead to a natural
interpretation of the spin incommensurability in terms of
antiferromagnetic domains separated by charge stripes oriented
along the principal axis.\cite{tran,zaanen,yama} However, at lower
doping, for $0.02<x<0.05$, in the so-called spin-glass regime,
where no superconductivity is observed, only two diagonal peaks
have been measured,\cite{NSSG} similar to the ones represented in
Fig.\ 1. Even though these peaks could still arise from diagonal
charge stripe formation, several arguments suggest that a spiral
picture in this regime is more likely, as discussed in Refs.
\onlinecite{antonio,nils,kotov}.

These observations stimulated further investigations on the
microscopic derivation of the $SO(3)$ NL$\sigma$ model
(\ref{eqkm}), which can allow for the determination of the various
microscopic parameters. Klee and Muramatsu\cite{klee} considered
the continuum field theory arising from a microscopic spin-fermion
model, where itinerant electrons are coupled via a Kondo-like
interaction to localized spins, described by the Heisenberg model.
The spin-fluctuations around the spiral configuration
(\ref{incomm}) are included by allowing the vectors $\bn_1$ and
$\bn_2$ to vary slowly on the lattice scale, and by adding a small
ferromagnetic frustration $\bL$, which is afterward integrated
out,
$$
\frac{\bf S}{S}= \frac{\bn+a\bL}{\sqrt{1+2a\bn\cdot
\bL+a^2\bL^2}}.
$$
By also taking into account the coupling to the fermions and
integrating them out, Klee and Muramatsu derived an effective
action for the spin field which is the $SO(3)$ NL$\sigma$ model
(\ref{eqkm}) with an additional term
\begin{eqnarray}
{\cal S}^{KM}=\int dt d^2{\bf x}  [\kappa_k(\partial_t \bn_k)^2-
p_{k\alpha}
\left(\partial_\alpha \bn_k\right)^2] \nonumber\\
\lb{eqkm2} -\int dt d^2{\bf x} s_\alpha \left(\bn_1\cdot
\partial_\alpha \bn_2\right).
\end{eqnarray}
In the above action both the exchange $J$ between the spins and
the fermionic susceptibilities contribute to the coupling
constants $\kappa$ and $p_{\alpha}$.  If only the Heisenberg
interaction between the spins is considered, then
$\kappa_1=\kappa_2\equiv\kappa=1/\{8Ja^2[2+\sum_\alpha\cos(Q_\alpha
a)]\}$, $\kappa_3\simeq0$, $p_{1\alpha}=p_{2\alpha}=
(JS^2/4)\cos(Q_\alpha a)$, $p_{3\alpha}\simeq0$ and
$s_\alpha=(JS^2/a)\sin(Q_\alpha a) $.   The subtle interplay
between holes and spins is represented in the last term of the
microscopically derived effective model (\ref{eqkm2}). Indeed,
since it is not positive definite, the weight of some field
configurations in the path integral will tend to infinity, hence
leading to instabilities.  In order to ensure the action to be at
a minimum, one should impose the condition that the {\em full}
coefficients $s_\alpha=\chi_\alpha-(JS^2/a)\sin(Q_\alpha a) $,
where $\chi_\alpha$ is the holes' contribution, must vanish. This
stability argument determines the spiral incommensurability
$Q_\alpha$ as a function of the microscopic parameters and the
doping concentration.\cite{klee}

A more general derivation of the stability condition has been
proposed recently by Hasselmann {\it et al.}\cite{nils} In this
approach, Eq.\ (\ref{eqkm2}) is considered as the continuum limit
of the Heisenberg model alone and the effect of doping is included
within a minimum coupling of the order parameter to a gauge field
$\bf B_\alpha$ representing the dipolar character of the hole
state, already emphasized by Shraiman and Siggia.
\cite{siggia881,siggia882,siggia89} It is then shown that the
stability condition
$$
p_{k\alpha}\partial_\alpha{\bf n}_k\cdot[{\bf B}_\alpha]_D\times
{\bf n}_k+s_\alpha{\bf n}_1\cdot
\partial_\alpha{\bf n}_2=0,
$$
where $[{\bf B}_\alpha]_D$ denotes the ordered fraction of
dipoles, relates the incommensurate vector ${\bf Q}$ to the hole
density, \cite{nils} in agreement with neutron scattering
measurements in lanthanum cuprates.\cite{NSSG} Moreover, the
dipolar frustration described within this minimal-coupling scheme
renormalizes the bare coefficients $p_{k\alpha}$ of Eq.\
(\ref{eqkm2}) leading the system toward a stable fixed point where
$p_{1\alpha}= p_{2\alpha}=p_{3\alpha} \equiv p_\alpha$,\cite{nils}
which corresponds to the model (\ref{eqkm1}) that we considered.
In this picture we can also determine the parameter
$\kappa=JS^2/4c^2$, where $c=\sqrt{c_{\parallel}c_{\perp}}\approx
2\sqrt{2}JSa$ and $(c_{\parallel}/c_\perp)^2=\cos(Qa)$.\cite{nils}
As a consequence, we can now apply our previous results to the
spin-glass phase of lanthanum cuprates.

\subsection{Inverse mobility in cuprates}

Until now we evaluated the inverse mobility of the defect without
specifying how this quantity can be accessed experimentally. As we
explained above, in lightly doped cuprates the holes act at the
same time as source and stabilizing mechanism of the dipolar
frustration. When topological defects are present in the spiral
spin texture, one could expect that the holes sit on top of the
defects (single vortex or vortex-antivortex pair) to minimize the
frustration.  Thus, the measured in-plane inverse mobility of the
holes would be described by Eq.\ (\ref{defmu}) with the damping
coefficient given by Eq.\ (\ref{barg2}). However, this scenario
should apply only for temperatures above $150$ K, because below
this temperature the experiments signal charge localization.

In the case of cuprates, the magnon temperature $E_c$ is the
antiferromagnetic coupling $J\approx 1200$ K measured at zero
doping. Actually, a lower value is expected, if one takes into
account the renormalization of the spin-spin interaction due to
the disorder introduced by hole doping and quantum effects. The
resulting inverse mobility as a function of temperature for the
case of a vortex-antivortex pair is reported in Fig.\ 6 for
several values of $E_c$ and $\alpha$. Observe that using
$a=3.8\AA$, as appropriate for cuprates, one obtains
$\mu_0^{-1}=0.46$ Vs/cm$^2$. Inspection of Fig.\ 6 shows that
already at $E_c=1000$ K the overall variation of $\mu^{-1}$
between 150 K and 300 K is of order of $0.05$ Vs/cm$^2$, as
observed experimentally. \cite{ando} Moreover, in the case of the
vortex-antivortex pair, an upper limit for the application of Eq.\
(\ref{mu}) is given by the temperature $T_v$ of the
vortex-antivortex unbinding. By estimating $T_v \sim \pi JS^2/2 $,
in analogy with the $XY$ model, we find that $T_v$ is of order of
400 K for the values of $E_c$ used in Fig.\ 6.

\begin{figure}[htb]
\begin{center}
\includegraphics[width=5.5cm,angle=-90]{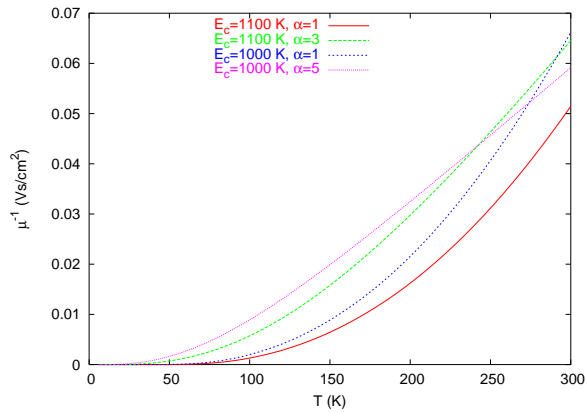}
\end{center}
\caption{(Color online) Contribution (\ref{mu}) of the motion of a
vortex-antivortex pair to the inverse mobility of the holes in
cuprates.}
\end{figure}

The inverse mobility of a single defect is shown in Fig.\ 7. The
overall variation of the inverse mobility turns out to be
quantitatively smaller in this case. As we explained in the
previous section, the main difference between the temperature
dependence of the inverse mobility obtained with the single vortex
or with the vortex-antivortex pair is that in the former case
$\mu^{-1}\propto T^2$ always, while in the latter case $\mu^{-1}$
evolves towards a linear behavior at a crossover temperature which
depends on $E_c$ and $\alpha$. As we discussed in Sec. II, the
presence of the two kinds of defects depends on their energy,
which scales as $E[\Psi_{1v}]\sim \ln \xi$ for the single vortex
and $E[\Psi_{2v}]\sim\ln d$ for the pair, where $\xi$ is the
correlation length and $d$ the distance between vortices,
respectively.  In the absence of disorder and at low temperatures,
one would expect $\xi$ to be finite, but still large enough to
prevent the formation of free defects below the crossover
temperature $T_v$, where pairs start to unbind.
\cite{kawamura1,southern,wintel2,caffarel} However, in Ref.
\onlinecite{nils} it has been argued that  disorder leads to a
strong reduction of the correlation length, and thus single
defects start to proliferate already at temperatures lower than
$T_v$. Comparison with recent resistivity data seems to support
this conclusion.  Indeed, studies performed by Ando {\it et
al.}\cite{ando1} for compounds in the spin-glass regime indicate
that the second derivative of the in-plane resistivity with
respect to the temperature is positive up to $\sim 300 K$,
implying that $\rho_{ab}\sim T^\eta$ with $\eta> 1$. Based on
previous experiments,\cite{ando} we interpreted  the resistivity
data for cuprates in terms of the dissipative motion of
vortex-antivortex pairs only, which gives rise to a linear
resistivity.\cite{juricic} However, in the light of these new
data, one could speculate about the coexistence of single-vortices
and vortex-antivortex pairs, which would consequently lead to a
power-law behavior of the resistivity with a more complicated
exponent, expected to be larger than one.

\begin{figure}[htb]
\begin{center}
\includegraphics[width=5.5cm,angle=-90]{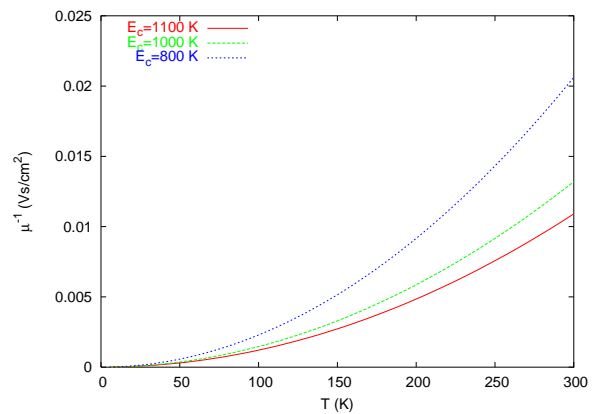}
\end{center}
\caption{(Color online) Contribution (\ref{mu}) of the motion of a
single-vortex defect to the inverse mobility of the holes in
cuprates.}
\end{figure}

\section{Conclusions}
We study here the properties  of frustrated Heisenberg spin
systems in which a noncollinear spin state is formed at low
temperatures. In the long-wavelength limit, the system is
described by the $SO(3)$ NL$\sigma$ model, and several differences
arise with respect to the usual $O(3)$ NL$\sigma$ model adopted to
describe collinear spin states. In particular, vortex-like
excitations play a crucial role in determining the
finite-temperature critical behavior.\cite{azaria} We concentrated
on the contribution of these topological defects to transport
properties. Our approach extends to a non-Abelian field theory the
well-known collective-coordinate method employed previously to
study the dissipative mechanism in one- and two-dimensional
systems. \cite{amir} We consider two kinds of topological defects:
a single vortex and a vortex-antivortex pair. We show that the
interaction between the defect and the spin waves is described by
a particle coupled to a bath of harmonic oscillators. The
scattering of the defect by the magnons leads to its dissipative
motion. We integrated out the bath and calculated the mobility of
the defect. Quite generally, its temperature dependence is
determined by the thermal activation of the magnons, which
vanishes at zero temperature and follows, at higher temperatures,
a power law whose exponent depends on the type of defect. In
particular, we find that it is linear for the vortex-antivortex
pair and quadratic for the single vortex. We apply the model to
describe transport in lightly doped lanthanum cuprates. Several
theoretical and experimental studies suggest that in these systems
a spiral state is formed at low
temperatures.\cite{nils,NSSG,juricic,ando} Our results for the
mobility indicate indeed that a possible mechanism for transport
in these materials, for 150K$<T<$400K, could be the dissipative
motion of an electrical charge attached to a single vortex or a
vortex-antivortex topological defect.

Although we  have applied the model to the particular case of
lightly doped cuprates, the approach presented here is quite
general, and can be employed to investigate the role of
topological defects in any frustrated spin system described by the
$SO(3)$ NL$\sigma$ model. As far as the spin-glass phase of
cuprate superconductors is concerned, the incommensurate peaks,
the value of the resistivity, as well as its linear temperature
dependence and anisotropy, might be explained within both the
spiral and the stripe model.\cite{nils,NSSG,juricic,ando} A
theoretical prediction that would discriminate between these two
scenarios, as well as its experimental realization, are
 still missing. This issue
is currently under investigation. \cite{barbosa}
\section{Acknowledgments}
We would like to acknowledge fruitful discussions with  N.\
Hasselmann, A.\ H.\ Castro Neto, A.\ V.\ Ferrer and M.\ B.\ Silva
Neto. This work was supported by the projects 620-62868.00 and
MaNEP (9,10,18) of the Swiss National Foundation for Scientific
Research. A.O.C. wishes to thank the partial support of Conselho
Nacional de Desenvolvimento Cient\'{\i}fico e Tecnol\'{o}gico
(CNPq) and Funda\c{c}\~{a}o de Amparo \`{a} Pesquisa no Estado de
S\~{a}o Paulo (FAPESP).

\appendix
\section{Energy of a single vortex and of a vortex-antivortex pair}
In this appendix we will calculate the energy of the topological
defects in the $SO(3)$ NL$\sigma$ model. Static solutions of the
model obey the Laplace equation
\begin{equation}\lb{laplace}
\nabla^2\Psi({\bf r})=0.
\end{equation}
A single-vortex solution centered at ${\bf R}=(X,Y)$
has the form
\begin{equation}\lb{sv}
\Psi_{1v}({\bf r},{\bf R})=\arctan\left(\frac{x-X}{y-Y}\right).
\end{equation}
Its energy is given by
$$
E[\Psi_{1v}]=\mathcal{N}c^2\int d^2{\bf r}
(\nabla\Psi_{1v})^2=2\pi\mathcal{N}c^2\ln\frac{\ell}{a}.
$$
A bound vortex-antivortex pair described by
\begin{equation}\lb{vapair}
\Psi_{2v}=\Psi_{1v}({\bf r},{\bf R}_1)-\Psi_{1v}({\bf r},{\bf R}_2),
\end{equation}
with $\Psi_{1v}$ given by Eq.\ (\ref{sv}), is also
a solution of Eq.\ (\ref{laplace}). The vortex-antivortex
defect can be written in a more compact form as
$$
\Psi_{2v}=\arctan\left\{\frac{[ {\bf d}\times ({\bf r}-{\bf
R})]_z}{({\bf r}-{\bf R})^2- {\bf d}^2/4}\right\},
$$
where the center of mass and relative coordinate,
respectively, are given by
$$
{\bf R}=\frac{1}{2}({\bf R}_1+{\bf R}_2),\qquad {\bf d}={\bf
R}_2-{\bf R}_1.
$$
In order to evaluate the energy of the vortex-antivortex pair
$$
E[\Psi_{2v}]=\mathcal{N}c^2\int d^2{\bf r}(\nabla\Psi_{2v})^2,
$$
we use Eq.\ (\ref{vapair}),
which yields
\begin{equation}\lb{I}
E[\Psi_{2v}]=\mathcal{N}c^2(I_{11}+I_{22}-2I_{12}),
\end{equation}
where
$$
I_{jj'}\equiv \int d^2{\bf r} (\nabla\Psi^{(j)})
(\nabla\Psi^{(j')}).
$$
and $\Psi^{(1,2)}\equiv \Psi_{1v}({\bf r},
{\bf R}_{(1,2)})$.
It is easy to show that $I_{11}$ and $I_{22}$ are
 equal,
\begin{equation}\lb{1122}
I_{11}=I_{22}=2\pi\ln\frac{\ell}{a}.
\end{equation}
The integral $I_{12}$ is
highly non-trivial. After some
calculations, it can be expressed in the form
 \begin{eqnarray} \lb {int12}
I_{12}&=&\frac{1}{2}\int d^2{\bf r} \frac{1}{({\bf r}-
{\bf R}_1)^2}+
\frac{1}{2}\int d^2{\bf r}
\frac{1}{({\bf r}-{\bf R}_2)^2}\nonumber\\
&-&\frac{1}{2}({\bf R}_1-{\bf R}_2)^2\int d^2{\bf r}
\frac{1}{({\bf r}-{\bf R}_1)^2({\bf r}-{\bf R}_2)^2}.
\end{eqnarray}
The first two integrals on the right hand side of
Eq.\ (\ref{int12}) are identical and equal to $I_{11}$,
whereas the last one must be evaluated
separately. Let us denote it as $I_{12}^{(3)}$. After
regularization, it becomes
$$
I_{12}^{(3)}=\int d^2{\bf r}
\frac{1}{[({\bf r}-{\bf R}_1)^2+a^2][({\bf r}-
{\bf R}_2)^2+a^2]}.
$$
By introducing new coordinates with ${\bf r}-{\bf
R}_2\rightarrow{\bf r}$, the last integral acquires the form
\begin{equation}  \lb {13}
I_{12}^{(3)}=\int d^2{\bf r}
\frac{1}{[{\bf r}^2+a^2][({\bf r}+{\bf d})^2+a^2]}.
\end{equation}
In order to simplify Eq.\ (\ref{13}),
 we introduce
 polar coordinates $x=r\cos\varphi,
y=r\sin\varphi$
$$
I_{12}^{(3)}=\int_0^\ell dr
\frac{r}{r^2+a^2}\int_0^{2\pi}
\frac{d\varphi}{r^2+2rd\cos(\varphi-\phi)+d^2+a^2},
$$
where $(d, \phi)$ are polar coordinates of ${\bf d}$.
 Integrating over the angle $\varphi$, we obtain
$$
I_{12}^{(3)}=\int_0^\ell dr
\frac{r}{r^2+a^2}\frac{2\pi}
{\sqrt{(r^2-d^2)^2+2a^2(r^2+d^2)+a^4}}.
$$
After substituting
$u=1/(r^2+a^2)$, the above integral reads
$$
I_{12}^{(3)}=\pi \int_{1/a^2}^{1/(\ell^2+a^2)}
\frac{du}{\sqrt{u^2 d^2(d^2+2a^2)-2ud^2+1}}.
$$
Straightforward calculations give
\begin{eqnarray} \lb {eq:intf}
I_{12}^{(3)}&=&\frac{\pi}{d\sqrt{d^2+2a^2}}\ln\left\{\frac{1}{4a^4}\left[d^3+\sqrt{d^6+8a^6}\right]\right.
\nonumber\\
&&\left. \left[d+\sqrt{d^2+2a^2}\right]\right\}
\approx \frac{4\pi}{d\sqrt{d^2+2a^2}}\ln\frac{d}{a}.
\end{eqnarray}
Substitution of Eq.\ (\ref {eq:intf})
into Eq.\ (\ref {int12}) yields
\begin{equation}
I_{12}=2\pi \left[\ln\frac{\ell}{a}-\frac{d}{\sqrt{d^2+2a^2}} \ln\frac{d}{a}\right].
\end{equation}
After inserting the last relation together with
Eq.\ (\ref{1122}) into Eq.\ (\ref {I}), we
finally obtain the energy of the vortex-antivortex pair
\begin{equation}
E[\Psi_{2v}]=\frac{4\pi d}{\sqrt{d^2+2a^2}}
\mathcal{N}c^2 \ln\frac{d}{a} \approx
4\pi \mathcal{N}c^2 \ln\frac{d}{a},
\end{equation}
which shows that the energy of the defect pair is
finite.
\section{Dynamics of the fluctuations around the
defect}

Using the identities
\begin{eqnarray}\lb{idsigma}
\sigma^a\sigma^b&=&\delta^{ab}+i\epsilon^{abc}
\sigma^c,\nonumber\\
\exp\left(\frac{i}{2}\vec{\sigma}\cdot\vec{\alpha}\right)
&=&\cos\frac{\alpha}{2}+
i\frac{\vec{\sigma}\cdot\vec{\alpha}}{\alpha}\sin
\frac{\alpha}{2},\nonumber
\end{eqnarray}
where $\alpha\equiv|\vec{\alpha}|$,
together with Eq.\ (\ref{gfull}), one can show that
\begin{eqnarray}\lb{derivg}
\partial_\mu g_s&=&\frac{i}{2}m^a\sigma^a g_s\partial_\mu\Psi_v,\\
\partial_\mu
g_\varepsilon&=&\frac{i}{2}\frac{\vec{\sigma}\cdot\vec{\varepsilon}}{\varepsilon}
\partial_\mu\varepsilon g_\varepsilon
+
\frac{i\sigma^a}{\varepsilon}\left[\partial_\mu\varepsilon^a-\frac
{\varepsilon^a(\vec{\varepsilon}\cdot\partial_\mu
\vec{\varepsilon})}{\varepsilon^2}\right]\sin\frac{\varepsilon}{2}.\nonumber
\end{eqnarray}
Here $\varepsilon$ stands for $|\vec{\varepsilon}|$.
By inserting Eqs.\ (\ref{derivg}) into the field
$A_\mu^a$ given by (\ref{A}), we obtain
\begin{equation}\lb{Amu}
A_\mu^a\equiv A_{\mu1}^a+A_{\mu2}^a+A_{\mu3}^a+A_{\mu4}^a,
\end{equation}
with
\begin{eqnarray}
A_{\mu1}^a&=&\frac{1}{4}{\rm tr}(\sigma^a
g_\varepsilon^{-1}g_s^{-1}\sigma^b g_s
g_\varepsilon)m^b\partial_\mu\Psi_v,\nonumber\\
A_{\mu2}^a&=&\frac{1}{2\varepsilon}{\rm tr}(\sigma^a
g_\varepsilon^{-1}\sigma^b)\partial_\mu\varepsilon^b\sin\frac{\varepsilon}{2},\nonumber\\
A_{\mu3}^a&=&-\frac{1}{2}{\rm tr}(\sigma^a
g_\varepsilon^{-1}\sigma^b)\frac{\varepsilon^b(\vec{\varepsilon}\cdot
\partial_\mu\vec{\varepsilon})}{\varepsilon^3}\sin\frac{\varepsilon}{2},\nonumber\\
A_{\mu4}^a&=&\frac{1}{4\varepsilon}{\rm tr}(\sigma^a
g_\varepsilon^{-1}\sigma^b\partial_\mu\varepsilon
g_\varepsilon)\varepsilon^b.\nonumber
\end{eqnarray}
The parameter $\varepsilon$ is small, $\varepsilon\ll 1$, because
$g_\varepsilon$ describes  fluctuations around the defect.
 Using the properties of the Pauli matrices, we find
$$
g_\varepsilon=1+\frac{i}{2}\vec{\sigma}\cdot
\vec{\varepsilon}
-\frac{1}{8}\varepsilon^2+\mathcal{O}(\varepsilon^3),
$$
which allows us to write
\begin{eqnarray}
 A_{\mu1}^a&=&\frac{1}{4}{\rm tr}\left\{\sigma^a\left(\cos\frac{\Psi_v}{2}
-i\vec{m}\cdot\vec{\sigma}\sin\frac{\Psi_v}{2}\right)
\sigma^b\right.\nonumber\\
&&\left. \left(\cos\frac{\Psi_v}{2}
+i\vec{m}\cdot\vec{\sigma}\sin\frac{\Psi_v}{2}\right)\right\}m^b\partial_\mu
\Psi_v\left(1-\frac{\varepsilon^2}{4}\right)\nonumber\\
&+& \frac{i}{8}{\rm tr}\left\{\sigma^a\left(\cos\frac{\Psi_v}{2}
-i\vec{m}\cdot\vec{\sigma}\sin\frac{\Psi_v}{2}\right)
\sigma^b \right.\nonumber\\
&&\left.\left(\cos\frac{\Psi_v}{2}
+i\vec{m}\cdot\vec{\sigma}\sin\frac{\Psi_v}{2}\right)(\vec{\sigma}
\cdot\vec{\varepsilon})\right\}m^b\partial_\mu
\Psi_v\nonumber\\
&-& \frac{i}{8}{\rm
tr}\left\{\sigma^a(\vec{\sigma}\cdot\vec{\varepsilon})\left(\cos\frac{\Psi_v}{2}
-i\vec{m}\cdot\vec{\sigma}\sin\frac{\Psi_v}{2}\right)
 \right.\nonumber\\
&&\left.\sigma^b\left(\cos\frac{\Psi_v}{2}
+i\vec{m}\cdot\vec{\sigma}\sin\frac{\Psi_v}{2}\right)\right\}m^b\partial_\mu\Psi_v\nonumber\\
&+&\frac{1}{16}{\rm tr}\left\{\sigma^a(\vec{\sigma}\cdot
\vec{\varepsilon})\left(\cos\frac{\Psi_v}{2}
-i\vec{m}\cdot\vec{\sigma}\sin\frac{\Psi_v}{2}\right)
\right.\nonumber\\
&&\left.\sigma^b\left(\cos\frac{\Psi_v}{2}
+i\vec{m}\cdot\vec{\sigma}\sin\frac{\Psi_v}{2}\right)
(\vec{\sigma}\cdot\vec{\varepsilon})\right\}m^b\partial_\mu\Psi_v.\nonumber
\end{eqnarray}
After some algebra, the last equation
acquires the form
\begin{eqnarray}\lb{Amu11}
A_{\mu1}^a&=&\frac{1}{2}m^a\partial_\mu\Psi_v
\left(1-\frac{\varepsilon^2}{4}\right)+
\frac{1}{2}\epsilon^{abc}\varepsilon^b m^c
\partial_\mu\Psi_v\nonumber\\
&+&\frac{1}{4}\varepsilon^a\varepsilon^b
m^b\partial_\mu\Psi_v-\frac{1}{8}\varepsilon^2m^a\partial_\mu
\Psi_v+\mathcal{O}(\varepsilon^3).
\end{eqnarray}
Analogously, one can show that
\begin{eqnarray}\lb{Amu12}
A_{\mu2}^a&=&\frac{1}{2}\partial_\mu\varepsilon^a+
\mathcal{O}(\varepsilon^3)\nonumber\\
A_{\mu3}^a&=&-\frac{\varepsilon^a
(\vec{\varepsilon}\cdot\partial_\mu\vec{\varepsilon})}
{\varepsilon^2}+\mathcal{O}(\varepsilon^3)\nonumber\\
A_{\mu4}^a&=&\frac{\varepsilon^a}{2\varepsilon}
\partial_\mu\varepsilon+\mathcal{O}(\varepsilon^3).
\end{eqnarray}
By substituting  Eqs.\ (\ref{Amu11}) and (\ref{Amu12}) into the field $A_\mu^a$ given by (\ref{Amu}),
and retaining terms up to the second order in $\varepsilon$,
we obtain
\begin{eqnarray}
A_\mu^a&=&\frac{1}{2}m^a\partial_\mu\Psi_v\left(1-
\frac{\varepsilon^2}{2}\right)+\frac{1}{4}\varepsilon^a
\varepsilon^b m^b\partial_\mu\Psi_v\nonumber\\
&+&\frac{1}{2}\epsilon^{abc}\varepsilon^b m^c\partial_\mu
\Psi_v+\frac{1}{2}\partial_\mu\varepsilon^a+
\frac{1}{4}\epsilon^{abc}\varepsilon^b
\partial_\mu\varepsilon^c,
\end{eqnarray}
and
\begin{eqnarray}
A_\mu^a A_\mu^a&=&\frac{1}{4}(\partial_\mu\Psi_v)^2+
\frac{1}{4}(\partial_\mu\vec{\varepsilon})^2+
\frac{1}{2}m^a\partial_\mu\varepsilon^a\partial_\mu\Psi_v
\nonumber\\
&+&\frac{1}{4}\epsilon^{abc}\partial_\mu\varepsilon^a
\varepsilon^b m^c\partial_\mu\Psi_v.
\end{eqnarray}

\section{Evaluation of the kernel}
In this appendix we will express the kernel $K$ defined by Eq.\
(\ref{defkernel})  as a functional integral. First, we divide the
time interval $[0,t]$ into $(m-1)$ subintervals of length
$\epsilon$, so $t=(m-1)\epsilon$, and use $(m-1)$ completeness
relations between the $(m-1)$ exponential functions,
\begin{eqnarray}
&&K({\bf x},\vec{\alpha}^*;{\bf y},\vec{\beta};t)
\equiv\langle{\bf x},\vec{\alpha}|e^{-\frac{i\hat{H}t}
{\hbar}}|{\bf y},\vec{\beta}\rangle\nonumber\\
&=& \int d^2{\bf x}_{m-1}...\int d^2{\bf x}_1\int
\frac{d^2\vec{\zeta}_{m-1}}{\pi^{2N}}...
\int\frac{d^2\vec{\zeta}_1}{\pi^{2N}}\nonumber\\
&&\langle{\bf x}_m\vec{\zeta}_m|
e^{-\frac{i\hat{H}\epsilon}{\hbar}}|{\bf x}_{m-1}
\vec{\zeta}_{m-1}\rangle\nonumber\\
&&\langle{\bf x}_{m-1}\vec{\zeta}_{m-1}|
e^{-\frac{i\hat{H}\epsilon}{\hbar}}|{\bf x}_{m-2}
\vec{\zeta}_{m-2}\rangle...\langle{\bf x}_1\vec{\zeta}_1|
e^{-\frac{i\hat{H}\epsilon}{\hbar}}|{\bf x}_0
\vec{\zeta}_0\rangle,\nonumber
\end{eqnarray}
where
\begin{eqnarray}\lb{C2}
&&{\bf x}_m\equiv{\bf x}, \quad {\bf x}_0\equiv {\bf y}
\nonumber\\
&&\vec{\zeta}_m\equiv\vec{\alpha},\quad
\vec{\zeta}_0\equiv\vec{\beta}.
\end{eqnarray}

By inserting $m$ completeness relations in the momentum
representation into Eq.\ (\ref{C2}), we obtain
\begin{eqnarray}\lb{K2}
&&K({\bf x},\vec{\alpha}^*;{\bf y},\vec{\beta};t)=
\left(\prod_{k=1}^{m-1}d^2{\bf x}_k\frac{d^2\vec{\zeta}_k}
{\pi^{2N}}\right)\left(\prod_{k=1}^m d^2{\bf P}_k\right)\nonumber\\
&&\prod_{k=1}^m\langle{\bf x}_k\vec{\zeta}_k|
e^{-\frac{i\hat{H}\epsilon}{\hbar}}|{\bf P}_k
\vec{\zeta}_{k-1}\rangle\langle{\bf P}_k| {\bf x}_{k-1}\rangle.
\end{eqnarray}
The matrix element $\langle{\bf x}_k\vec{\zeta}_k|
e^{-\frac{i\hat{H}\epsilon}{\hbar}}|{\bf P}_k
\vec{\zeta}_{k-1}\rangle$, $(k=1,...,m)$, can be evaluated using
that $\epsilon\ll t$. It reads
\begin{eqnarray}\lb{melement}
&&\langle{\bf x}_k\vec{\zeta}_k|
e^{-\frac{i\hat{H}\epsilon}{\hbar}}|{\bf P}_k
\vec{\zeta}_{k-1}\rangle=\langle{\bf x}_k| {\bf P}_k\rangle
\langle\vec{\zeta}_k|\vec{\zeta}_{k-1}\rangle
\nonumber\\
&&\exp\left[-\frac{i\epsilon}{\hbar} H({\bf x}_k,{\bf
P}_k;\vec{\zeta}_k^*,\vec{\zeta}_{k-1})\right],
\end{eqnarray}
where
\begin{eqnarray}
H({\bf x}_k,{\bf P}_k;\vec{\zeta}_k^*,
\vec{\zeta}_{k-1})\equiv
\frac{\langle{\bf x}_k\vec{\zeta}_k|
\hat{H}|{\bf P}_k\vec{\zeta}_{k-1}\rangle}
{\langle{\bf x}_k|{\bf P}_k\rangle
\langle\vec{\zeta}_k|\vec{\zeta}_{k-1}\rangle}.\nonumber
\end{eqnarray}
Substituting the matrix element given by
Eq.\  (\ref{melement}) into Eq.\ (\ref{K2}), recalling that
$$
\langle{\bf P}|{\bf x}\rangle=\frac{1}{2\pi\hbar}
\exp\left(-\frac{i}{\hbar}{\bf P}\cdot{\bf x}\right),
$$
and using the properties of the overlap of
coherent states, we find that the kernel acquires the form
\begin{eqnarray}\lb{K3}
&&K({\bf x},\vec{\alpha}^*;{\bf y},\vec{\beta};t)=
\int\frac{d^2{\bf P}_m}{2\pi\hbar}
\left(\prod_{l=1}^{m-1}\int
\frac{d^2{\bf x}_l d^2{\bf P}_l}{2\pi\hbar}
\frac{d^2{\vec \zeta}_l}{\pi^{2N}}\right)\nonumber\\
&&\exp\left\{\sum_{k=1}^m
\frac{1}{2}\left[
\vec{\zeta}_{k-1}\cdot(\vec{\zeta}^*_k-
\vec{\zeta}^*_{k-1})-
\vec{\zeta}^*_k\cdot(\vec{\zeta}_k-
\vec{\zeta}_{k-1})\right]\right.\nonumber\\
&+&\left.\frac{i}{\hbar}\left[{\bf P}_k ({\bf x}_k-{\bf
x}_{k-1})-\epsilon H({\bf x}_k,{\bf P}_k;\vec{\zeta}_k^*,
\vec{\zeta}_{k-1})\right]\right\}.
\end{eqnarray}

In order to integrate over the momenta ${\bf P}_k$, we have
to explicitly calculate
$H({\bf x}_k,{\bf p}_k;\vec{\zeta}_k^*,
\vec{\zeta}_{k-1})$. Using the coherent state
 representation for the bath of harmonic oscillators, we
find
\begin{eqnarray}
&&H({\bf x}_k,{\bf P}_k;\vec{\zeta}_k^*,
\vec{\zeta}_{k-1})=\frac{{\bf P}_k^2}{2M}+\sum_{nm,i=a,b}
\hbar\omega_{nm}\zeta_{nm,i}^{k*}\zeta_{nm,i}^{k-1}
\nonumber\\
&-&\frac{{\bf P}_k}{M}\sum_{nm,pq}
\left[{\bf D}_{nm,pq}\zeta_{pq,a}^{k*}
\zeta_{nm,a}^{k-1}-{\bf D}_{pq,nm}
\zeta_{nm,b}^{k-1}\zeta_{pq,b}^{k*}\right],\nonumber
\end{eqnarray}
which, after insertion into Eq.\ (\ref{K3}) and
integration over the momenta ${\bf P}_k$, yields
\begin{eqnarray}\lb{K4}
&&K({\bf x},\vec{\alpha}^*;{\bf y},\vec{\beta};t)=
\left(\frac{M}{2i\pi\hbar}\right)^{m/2}
\left(\prod_{l=1}^{m-1}\frac{d^2{\bf x}_l}
{2\pi\hbar}\frac{d^2{\vec \zeta}_l}{\pi^{2N}}\right)
\\
&&\exp\left\{\sum_{k=1}^m\frac{1}{2}\left[
\vec{\zeta}_{k-1}(\vec{\zeta}^*_k-
\vec{\zeta}^*_{k-1})-
\vec{\zeta}^*_k(\vec{\zeta}_k-
\vec{\zeta}_{k-1})\right]\right.\nonumber\\
&&\left.-\frac{i\epsilon}{\hbar}\sum_{nm,i}\hbar\omega_{nm}
\zeta_{nm,i}^{k*}\zeta_{nm,i}^{k-1}+
\frac{iM\epsilon}{2\hbar}\left({\bf h}_k+ \frac{{\bf x}_k-{\bf
x}_{k-1}} {\epsilon}\right)^2\right\},\nonumber
\end{eqnarray}
with
$$
{\bf h}_k\equiv\frac{\hbar}{M}\sum_{nm,pq} \left({\bf
D}_{nm,pq}\zeta_{pq,a}^{k*}\zeta_{nm,a}^{k-1} -{\bf
D}_{pq,nm}\zeta_{nm,b}^{k-1}\zeta_{pq,b}^{k*}\right).
$$
Now, we consider the continuum limit, $\epsilon\to 0$, of the last
equation. Using the boundary conditions (\ref{C2}), the first term
in the exponential in (\ref{K4}) becomes
\begin{eqnarray}
&&\exp\left\{\sum_{k=1}^m\frac{1}{2}\left[
\vec{\zeta}_{k-1}\cdot(\vec{\zeta}^*_k-
\vec{\zeta}^*_{k-1})-
\vec{\zeta}^*_k\cdot(\vec{\zeta}_k-
\vec{\zeta}_{k-1})\right]\right\}\nonumber\\
&=&\exp\left\{\frac{1}{2}\sum_{k=0}^{m-1}
\vec{\zeta}_k\cdot(\vec{\zeta}_{k+1}^*-
\vec{\zeta}_k^*)-\frac{1}{2}\sum_{k=1}^m
\vec{\zeta}_k^*\cdot(\vec{\zeta}_k-
\vec{\zeta}_{k-1})\right\}\nonumber\\
&=& \exp\left\{\frac{1}{2}[\vec{\zeta}_0
\cdot(\vec{\zeta}_1^*-\vec{\zeta}_0^*)-\vec{\zeta}_m^*
\cdot(\vec{\zeta}_m-\vec{\zeta}_{m-1})]\right\}\nonumber\\
&&\exp\left\{\frac{1}{2}\sum_{k=1}^{m-1}[\vec{\zeta}_k\cdot(\vec{\zeta}_{k+1}^*-
\vec{\zeta}_k^*)-\vec{\zeta}_k^*\cdot(\vec{\zeta}_k-
\vec{\zeta}_{k-1})]\right\}\nonumber\\
&\rightarrow&\exp\left(-\frac{1}{2}|\vec{\beta}|^2-
\frac{1}{2}|\vec{\alpha}|^2\right)
\exp\left\{\frac{1}{2}\vec{\beta}\cdot\vec{\zeta}^*(0)+
\frac{1}{2}\vec{\alpha}^*\cdot\vec{\zeta}(t)\right\}
\nonumber\\
&&\exp\left\{\frac{1}{2}\int_0^t dt'
(\vec{\zeta}\cdot\dot{\vec{\zeta}^*}-
\vec{\zeta}^*\cdot\dot{\vec{\zeta}})\right\}.\nonumber
\end{eqnarray}
 The other terms in Eq.\ (\ref{K4}) can be
trivially written in the continuum limit,
yielding then Eq.\ (\ref{kernel}).
\section{Initial density matrix for the bath in the
coherent state representation}

It remains to evaluate the matrix elements of the
initial density matrix for the bath in the coherent state
representation
$$
\rho_B({\vec \beta}^*,{\vec \beta}',0)=\langle
{\vec \beta|\hat{\rho}_B(0)|{\vec \beta}'}\rangle
$$
with
$$
\hat{\rho}_B(0)=\frac{1}{Z}
\prod_{pq}e^{-U\omega_{pq}(\hat{a}_{pq}^\dagger
\hat{a}_{pq}+\hat{b}_{pq}^\dagger \hat{b}_{pq})},
$$
where the partition function reads
$$
Z\equiv {\rm tr}\left[e^{-U\sum_{pq}\omega_{pq}
(\hat{a}_{pq}^\dagger\hat{a}_{pq}+\hat{b}_{pq}^\dagger
\hat{b}_{pq})}\right]
=\prod_{pq}\left(1-e^{-\frac{\hbar\omega_{pq}}
{k_BT}}\right)^{-2}.
$$
Since the baths $a$ and $b$ are not coupled, the total
density matrix is the product
$$
\rho_B({\vec \beta}^*,{\vec \beta}',0)=
\rho_{B,a}({\vec \beta}_a^*,{\vec \beta}'_a,0)
\rho_{B,b}({\vec \beta}_b^*,{\vec \beta}'_b,0)
$$
of the density matrices $\rho_{B,a}$ and $\rho_{B,a}$
for the baths $a$ and $b$, respectively, given by
$$
\rho_{B,a}({\vec \beta}_a^*,{\vec \beta}'_a,0)= \frac{1}{Z_a}
\prod_{pq}\langle\beta_{pq,a}|e^{-U\omega_{pq} \hat{a}_{pq}
^\dagger \hat{a}_{pq}}|\beta'_{pq,a}\rangle
$$
for the bath $a$, and analogously for the bath $b$, with the
partition function
$$
Z_b=Z_a={\rm tr}\left[e^{-U\sum_{pq}\omega_{pq}
(\hat{a}_{pq}^\dagger\hat{a}_{pq})}\right]
=\prod_{pq}\left(1-e^{-\frac{\hbar\omega_{pq}}
{k_BT}}\right)^{-1}.
$$
We shall evaluate the previous matrix element by inserting two
unity operators in the occupation number representation for the
bath $a$ (in the following, to simplify notation we omit the index
$a$ in the ${\vec \beta}$, ${\vec \beta}'$ and ${\bar n}$)
\begin{eqnarray}\lb{rhob1}
\rho_{B,a}({\vec \beta}^*,{\vec \beta}',0)&=&\frac{1}{Z_a}
\prod_{pq}\langle\beta_{pq}|e^{-U\omega_{pq} \hat{a}_{pq} ^\dagger
\hat{a}_{pq}}|\beta'_{pq}\rangle
\\
&=&\frac{1}{Z_a}\prod_{pq}\sum_{n_{pq},n'_{pq}}
\langle\beta_{pq}|n_{pq}\rangle\nonumber\\
&&\langle n_{pq} |e^{-U\omega_{pq}\hat{a}_{pq}^\dagger
\hat{a}_{pq}} |n'_{pq}\rangle\langle n'_{pq}|\beta'_{pq}
\rangle.\nonumber
\end{eqnarray}
Now, we use the scalar product of the states which
define the coherent state and occupation number
representations
$$
\langle\beta_{pq}|n_{pq}\rangle=\frac{(\beta^*_{pq})^
{n_{pq}}}{\sqrt{n_{pq}!}}e^{-\frac{|\beta_{pq}|^2}{2}}.
$$
Substituting the above expression into Eq.\ (\ref{rhob1}),
we obtain
\begin{eqnarray}\lb{cohbdm}
\rho_{B,a}({\vec \beta}^*,{\vec \beta}',0)&=&\frac{1}{Z_a}
\prod_{nm}\exp\left(-\frac{|\beta_{pq}|^2}{2}-
\frac{|\beta'_{pq}|^2}{2}\right)
\nonumber \\
&&\sum_{n_{pq}}\frac{1}{n_{pq}!}\left\{\exp\left(-
\frac{\hbar\omega_{pq}}{k_BT}\right)\beta^*_{pq}
\beta'_{pq}\right\}^{n_{pq}}
\nonumber \\
&=& \frac{1}{Z_a}\prod_{pq}\exp\left(-\frac{|\beta_{pq}|^2}{2}-
\frac{|\beta'_{pq}|^2}{2}\right)\nonumber\\
&&\exp\left[\beta^*_{pq}
\beta'_{pq}\exp(-U\omega_{pq})\right].
\end{eqnarray}
\section{Equations of motion}
Our next step is to solve
equations of motion (\ref{eqm1}). In order to achieve this aim,
we introduce the ansatz
\begin{eqnarray}
\lb{alpha1} \zeta_{nm}(\tau)&=&e^{-i\omega_{nm}\tau}
[\beta_{nm}+\sum_{kl\neq nm}W_{nm,kl}(\tau)\beta_{kl}],
\nonumber \\
\zeta^*_{nm}(\tau)&=&e^{i\omega_{nm}\tau}
[\alpha^*_{nm}e^{-i\omega_{nm}t}\nonumber\\
&+&\sum_{kl\neq nm}\tilde{W}_{nm,kl}(\tau)e^{-i\omega_{kl}t}
\alpha^*_{kl}],
\end{eqnarray}
where the functionals $W$ and $\tilde{W}$ will be determined
from the equations of motion and
$0\leq\tau\leq t$. By substituting the first time
derivative of Eqs.\ (\ref{alpha1})  into
the equations of motion (\ref{eqm1}), we obtain
the expressions which determine the time evolution of the
functionals $W$ and $\tilde{W}$
$$
\dot{W}_{nm,kl}=W^0_{nm,kl}+\sum_{pq}W^0_{nm,pq}W_{pq,kl},
$$
$$
\dot{\tilde{W}}_{nm,kl}=-\tilde{W}^0_{nm,kl}-
\sum_{pq}\tilde{W}^0_{nm,pq}\tilde{W}_{pq,kl},
$$
with
\begin{eqnarray}
\lb{defW0} W^0_{nm,kl}[{\bf x},\tau]&=&i\dot{{\bf x}}{\bf D}_
{kl,nm}e^{i(\omega_{nm}-\omega_{kl})\tau},
\nonumber\\
\tilde{W}^0_{nm,kl}[{\bf x},\tau]&=&i\dot{{\bf x}} {\bf
D}_{nm,kl}e^{-i(\omega_{nm}-\omega_{kl})\tau}.
\end{eqnarray}
Notice that $\tilde{W}^0_{nm,kl}=-W^{0*}_{nm,kl}$.
Because $W$ and $\tilde{W}$ must satisfy the boundary conditions
$W({\bf x},0)=0$ and $\tilde{W}({\bf x},t)=0$, we have
\begin{eqnarray}\lb{W1}
W_{nm,kl}[{\bf x},\tau]&=&\int_0^\tau dt'
W^0_{nm,kl}[{\bf x},t']
\nonumber\\
&+&\sum_{pq}\int_0^
\tau dt'W^0_{nm,pq}[{\bf x},t'] W_{pq,kl}[{\bf x},t'],
\nonumber\\
\tilde{W}_{nm,kl}[{\bf x},\tau]&=&\int_\tau^t dt'
\tilde{W}^0_{nm,kl}[{\bf x},t']
\nonumber \\
&+&\sum_{pq}\int_\tau^t dt'
\tilde{W}^0_{nm,pq}[{\bf x},t'] \tilde{W}_{pq,kl}
[{\bf x},t'],\nonumber\\
\end{eqnarray}
which, using the Born approximation, acquires the form
\begin{eqnarray}
&&W_{nm,kl}[{\bf x},\tau]=\int_0^\tau dt'W_{nm,kl}^0[{\bf x},t']
\nonumber\\
&+&\sum_{pq}\int_0^\tau dt' W^0_{nm,pq}[{\bf x},t']
\int_0^{t'}
dt''W^0_{pq,kl}[{\bf x},t'],\nonumber
\end{eqnarray}
\begin{eqnarray}\lb{BtW}
&&\tilde{W}_{nm,kl}[{\bf x},\tau]=\int_\tau^t dt'
\tilde{W}_{nm,kl}^0[{\bf x},t']\nonumber\\
&+&
\sum_{pq}\int_\tau^t dt' \tilde{W}^0_{nm,pq}[{\bf x},t']
\int_{t'}^{t}
dt''\tilde{W}^0_{pq,kl}[{\bf x},t'].
\end{eqnarray}
The functions $\gamma$ appearing in
Eq.\ (\ref{influence5}) obey the equations of motion
(\ref{eqm1}) with the boundary conditions (\ref{ic2}).
We solve them by introducing the ansatz
\begin{eqnarray}\lb{gamma1}
\gamma_{nm}(\tau)&=&e^{-i\omega_{nm}\tau}[\alpha_{nm}
e^{i\omega_{nm}t}
\nonumber\\
&+&\sum_{kl\neq nm}
\bar{W}_{nm,kl}(\tau)\alpha_{kl}e^{i\omega_{kl}t}],
\nonumber\\
\gamma^*_{nm}(\tau)&=&e^{i\omega_{nm}\tau}[\beta'^*_{nm}
+\sum_{kl\neq nm} \tilde{\bar{W}}_{nm,kl}(\tau)\beta'^*_{kl}],
\end{eqnarray}
with the conditions
$\bar{W}(t)=0$ and $\tilde{\bar{W}}(0)=0$.
By inserting this ansatz  into the
corresponding equations of motion, we find, after some algebra,
\begin{eqnarray}
\tilde{\bar{W}}_{nm,kl}[{\bf x},\tau]&=&\int_0^\tau dt'
\tilde{\bar{W}}^0_{nm,kl}[{\bf x},t']
\nonumber \\
&+&\sum_{pq}\int_0^
\tau dt'\tilde{\bar{W}}^0_{nm,pq}[{\bf x},t']
\tilde{\bar{W}}_{pq,kl}[{\bf x},t'],
\nonumber\\
\bar{W}_{nm,kl}[{\bf x},\tau]&=&\int_\tau^t dt'
\bar{W}^0_{nm,kl}[{\bf x},t']
\nonumber \\
&+&\sum_{pq}\int_\tau^t dt'
\bar{W}^0_{nm,pq}[{\bf x},t'] \bar{W}_{pq,kl}[{\bf x},t'],\nonumber\\
\end{eqnarray}
with
$$
\tilde{\bar{W}}^0_{nm,kl}=W^{0*}_{nm,kl},\qquad
\bar{W}^0_{nm,kl}=\tilde{W}^{0*}_{nm,kl}.
$$
The boundary values of the functionals obey the relations
$$
\bar{W}_{nm,kl}[{\bf x},0]=\tilde{W}^*_{nm,kl}[{\bf x},0],
$$
\begin{equation}\lb{tildewwstar}
\tilde{\bar{W}}_{nm,kl}[{\bf x},t]=W^*_{nm,kl}[{\bf x},t],
\end{equation}
which will be used later. From
Eqs.\ (\ref{alpha1}) and (\ref{gamma1}) the boundary terms read
$$
\zeta_{nm}(t)=\beta_{nm} e^{-i\omega_{nm}t}+\sum W_{nm,kl}({\bf
x},t)e^{-i\omega_{nm}t}\beta_{kl},
$$
$$
\zeta_{nm}^*(0)=\alpha_{nm}^* e^{-i\omega_{nm}t}+\sum
\tilde{W}_{nm,kl}({\bf x},0)e^{-i\omega_{kl}t}\alpha_{kl}^*,
$$
$$
\gamma_{nm}(0)=\alpha_{nm} e^{i\omega_{nm}t}+\sum
\bar{W}_{nm,kl}({\bf y},0)e^{i\omega_{kl}t}\alpha_{kl},
$$
\begin{equation}\lb{gammat}
\gamma_{nm}^*(t)=\beta'^*_{nm} e^{i\omega_{nm}t}+\sum
\tilde{\bar{W}}_{nm,kl}({\bf y},t)e^{i\omega_{nm}t}\beta'^*_{kl}.
\end{equation}
\begin{widetext}
\section{Evaluation of $\Gamma_{nm,nm}$}

In this appendix we evaluate  the diagonal elements
of the matrix $\Gamma=\Gamma^a+\Gamma^b$, where the
elements $\Gamma^a_{nm,nm}$ are given by
\begin{eqnarray}\lb{diaggamma}
\Gamma^a_{nm,nm}&=&\frac{1}{2}\left[W_{nm,nm}[{\bf x},t]
+\tilde{W}_{nm,nm}[{\bf x},0]+
\tilde{W}^*_{nm,nm}
 [{\bf y},0]+W^*_{nm,nm}[{\bf y},t]\right]+
\frac{1}{4}\sum_{pq}\left[\tilde{W}_{nm,pq}
[{\bf x},0]+ W_{pq,nm}[{\bf x},t] \right]\nonumber\\
&&\left[\tilde{W}^*_{nm,pq}[{\bf y},0]+
W^*_{pq,nm}[{\bf y},t]\right],
\end{eqnarray}
and the ones of matrix $\Gamma^b$ are obtained from the latter by
the substitution ${\bf D}_{nm,kl}\rightarrow -{\bf
D}_{kl,nm}=-{\bf D}^*_{nm,kl}$. Using the Born approximation for
the functionals $W$ and $\tilde{W}$ given by Eq.\ (\ref{BtW}), and
the form of the functionals $W^0$ and $\tilde{W}^0$, defined after
Eq.\ (\ref{alpha1}), we find
\begin{eqnarray} \lb{diaggamma5}
W_{nm,nm}[{\bf x},t]&=&-\sum_{\mu,\nu,pq}\int_0^t dt'
\int_0^t dt'' \theta(t'-t'')\dot{x}^\mu(t')\dot{x}^\nu
(t'') D^{\mu*}_{nm,pq}D^\nu_{nm,pq}
e^{i(\omega_{nm}-\omega_{pq})(t'-t'')}\nonumber\\
\tilde{W}_{nm,nm}[{\bf x},0]&=&-\sum_{\mu,\nu,pq}\int_0^t dt'
\int_0^t dt'' \theta(t''-t')\dot{x}^\mu(t')
\dot{x}^\nu(t'')D^\mu_{nm,pq}D^{\nu*}_{nm,pq}e^{i(\omega_{nm}-
\omega_{pq})(t''-t')}=W_{nm,nm}[{\bf x},t].\nonumber\\
\end{eqnarray}
Using Eq.\ (\ref{W1}), as well as its
complex conjugate evaluated at ${\bf y}$ and
retaining only the terms quadratic in the coupling
constants, we can write the last term in
Eq.\ (\ref{diaggamma}) as
\begin{eqnarray} \lb{diaggamma6}
&&\frac{1}{4}\sum_{pq}[\tilde{W}_{nm,pq}({\bf x},0)
+W_{pq,nm}({\bf x},t)][\tilde{W}^*_{nm,pq}
({\bf y},0)
+W^*_{pq,nm}({\bf y},t)]\simeq \frac{1}{4}\int_0^t dt'
\int_0^t dt''\left[\tilde{W}^0_{nm,pq}[{\bf x},t']
+W^{0}_{pq,nm}[{\bf x},t']\right]\nonumber\\
&&\left[\tilde{W}^{0*}_{nm,pq}
[{\bf y},t'']+W^{0*}_{pq,nm}[{\bf y},t'']\right]
=\sum_{\mu,\nu,pq}\int_0^t dt'\int_0^t dt''\dot{x}^\mu(t')
\dot{y}^\nu(t'')D^\mu_{nm,pq}D^{\nu*}_{nm,pq}
e^{i(\omega_{pq}-\omega_{nm})(t'-t'')}\nonumber\\
&=&\int_0^t dt'\int_0^t dt''\theta(t'-t'')
D^\mu_{nm,pq}D^{\nu*}_{nm,pq}\left(\dot{x}^\mu(t')
\dot{y}^\nu(t'')
e^{i(\omega_{pq}-\omega_{nm})(t'-t'')}+\dot{x}^\mu(t'')
\dot{y}^\nu(t') e^{-i(\omega_{pq}-\omega_{nm})(t'-t'')}\right).
\end{eqnarray}

Substituting Eqs.\ (\ref{diaggamma5}) and (\ref{diaggamma6})
into Eq.\ (\ref{diaggamma}), we obtain the diagonal elements
of the matrix $\Gamma^a$, in the lowest order in the coupling constants
\begin{eqnarray}
\Gamma^a_{nm,nm}&=&-\sum_{\mu,\nu,pq}\int_0^t dt'\int_0^t dt''\theta(t'-t'')[{\dot x}^\mu(t')-{\dot y}^\mu(t')]
\left[{\dot x}^\nu(t'') D^{\mu*}_{nm,pq}D^\nu_{nm,pq}
e^{i(\omega_{nm}-\omega_{pq})(t'-t'')}\right.\nonumber\\
&-&\left.{\dot y}^\nu(t'')D^{\mu}_{nm,pq}D^{\nu*}_{nm,pq}
e^{-i(\omega_{nm}-\omega_{pq})(t'-t'')}\right],
\end{eqnarray}
yielding the diagonal elements of the matrix $\Gamma$
\begin{eqnarray}\lb{diaggamma1}
\Gamma_{nm,nm}&=&-\sum_{\mu,\nu,pq}\int_0^t dt'\int_0^t dt''\theta(t'-t'')\left(D^{\mu}_{nm,pq}D^{\nu*}_{nm,pq}+
D^{\mu*}_{nm,pq}D^\nu_{nm,pq}\right)
[{\dot x}^\mu(t')-{\dot y}^\mu(t')]\nonumber\\
&&\left[{\dot x}^\nu(t'')
e^{i(\omega_{nm}-\omega_{pq})(t'-t'')}-{\dot y}^\nu(t'')
e^{-i(\omega_{nm}-\omega_{pq})(t'-t'')}\right].
\end{eqnarray}
\section{Evaluation of the coupling constants}

In this appendix we calculate the coupling constants
\begin{equation}
{\bf G}^*_{km,k'l}=\int d^2{\bf r}\eta_{k'l}^*\nabla \eta_{km},
\end{equation}
where the wave functions are given by
\begin{equation}
\lb{etaf} \eta_{km}=\sqrt{\frac{k}{2\ell}}
\left[H_m^{(1)}(kr)+e^{-2i\delta_m}H_m^{(2)}(kr)\right]
e^{im\vartheta}, \qquad (m>0).
\end{equation}
By expressing the gradient operator in polar coordinates  we find
that
\begin{eqnarray}\lb{intcc}
G^{x*}_{km,k'l}&=& \pi \delta_{m-l,1}
\int_0^\infty dr r [R_l(k'r)S_m(kr)-mR_l(k'r)F_m(kr)]\nonumber\\
&+&\pi \delta_{m-l,-1}
\int_0^\infty dr r [R_l(k'r)S_m(kr)+mR_l(k'r)F_m(kr)]
,\nonumber\\
G^{y*}_{km,k'l}&=& i\pi \delta_{m-l,1}
\int_0^\infty dr r [R_l(k'r)S_m(kr)+mR_l(k'r)F_m(kr)]\nonumber\\
&-&i\pi \delta_{m-l,-1} \int_0^\infty dr r
[R_l(k'r)S_m(kr)-mR_l(k'r)F_m(kr)], \lb{gg}
\end{eqnarray}
where
\begin{eqnarray}
R_l(kr) &=& \sqrt\frac{k}{2 \ell}[H^{(2)}_l(kr)+e^{2i\delta_l}H^{(1)}_l(kr)],\\
S_l(kr) &=& \frac{k}{2}\sqrt\frac{k}{2
\ell}[H^{(1)}_{l-1}(kr)+e^{-2i\delta_l}H^{(2)}_{l-1}(kr)-H^{(1)}_{l+1}(kr)+
\nonumber\\
&+&e^{-2i\delta_l}H^{(2)}_{l+1}(kr)],\\
F_l(kr) &=& \frac{l}{r}\sqrt\frac{k}{2
\ell}[H^{(1)}_l(kr)+e^{-2i\delta_l} H^{(2)}_l(kr)].
\end{eqnarray}
In evaluating the expression (\ref{gg}) one can use the asymptotic
form of the Hankel functions
\begin{eqnarray}
&&H_m^{(1)}(kr)=\sqrt{\frac{2}{\pi k r}}e^{ikr}
e^{-i\frac{\pi}{2}(m+1/2)},\nonumber\\
&&H_m^{(2)}(kr)=\sqrt{\frac{2}{\pi k r}}e^{-ikr}
e^{i\frac{\pi}{2}(m+1/2)}.\nonumber
\end{eqnarray}
However, the terms coming from $F_m(kr)$ would then be divergent
at $r=0$. This divergence is an artifact of approximating up to
$r=0$ the true solution of the scattering problem
(\ref{schroedinger}) with the functions (\ref{etaf}). At small $r$
indeed a better approximation for the radial part of the functions
$\eta_{km}$ is provided by the Bessel functions $J_m(kr)$, which
are regular at $r=0$. Then we find that
\begin{eqnarray}
\int_0^\infty dr r H_\nu^{(1,2)}(k'r) H_{\mu}^{(1,2)}(kr)
&=&\frac{2e^{\pm i\pi(\mu+\nu+1)/2}}{\pi\sqrt{kk'}} \left(\pi
\delta(k+ k')\pm \mathcal{P}\frac{1}{k+ k'}
\right),\nonumber\\
\int_0^\infty dr r H_\nu^{(1)}(k'r) H_{\mu}^{(2)}(kr)
&=&\frac{2e^{ i\pi(\mu-\nu)/2}}{\pi\sqrt{kk'}} \left(\pi \delta(k-
k')+ \mathcal{P}\frac{1}{k- k'}
\right),\nonumber\\
\int_0^\infty dr
J_{m+1}(k'r)J_{m}(kr)&=&\frac{k^m}{(k')^{m+1}}\Theta(k'-k)
+\frac{1}{2k}\delta(k-k').
\end{eqnarray}
Because the terms in $S(\omega,\omega')$ proportional to
$\delta(\omega\pm\omega')$ do not contribute to the damping matrix
(\ref{defdampmatrix}), we can directly discard them from the
definition of the coupling constants, which are then given by
\begin{eqnarray}\lb{explcc}
G_{km,k'l}^{x*}&=&\frac{k}{\ell}\delta_{l,m\pm1}
\left\{i\Lambda_{lm}^{(1)}\mathcal{P}
\frac{1}{k+k'}-\Lambda_{lm}^{(2)}\mathcal{P}\frac{1}{k-k'}\right\}
+\delta_{l,m-1}\frac{m\pi\sqrt{kk'}}{4\ell} \left[
\frac{(k')^m}{k^{m+1}}\Theta(k-k')
+\frac{1}{2k}\delta(k-k')\right]\nonumber\\
&-&\delta_{l,m+1}\frac{m\pi\sqrt{kk'}}{4\ell} \left[
\frac{k^m}{(k')^{m+1}}\Theta(k'-k)
+\frac{1}{2k}\delta(k'-k)\right],\nonumber\\
G_{km,k'l}^{y*}&=&i\frac{k}{\ell}(\delta_{l,m-1}- \delta_{l,m+1})
\left\{i\Lambda_{lm}^{(1)}\mathcal{P}
\frac{1}{k+k'}-\Lambda_{lm}^{(2)}\mathcal{P}\frac{1}{k-k'}\right\}
+i\delta_{l,m-1}\frac{m\pi\sqrt{kk'}}{4\ell} \left[
\frac{(k')^m}{k^{m+1}}\Theta(k-k')
+\frac{1}{2k}\delta(k-k')\right]\nonumber\\
&+&i\delta_{l,m+1}\frac{m\pi\sqrt{kk'}}{4\ell} \left[
\frac{k^m}{(k')^{m+1}}\Theta(k'-k)
+\frac{1}{2k}\delta(k'-k)\right],
\end{eqnarray}
where
$$
\Lambda^{(1)}_{lm}=-e^{i\pi(l+m)/2}e^{-2i\delta_m}+
e^{-i\pi(l+m)/2}e^{2i\delta_l},\qquad
\Lambda^{(2)}_{lm}=e^{i\pi(l-m)/2}+e^{-i\pi(l-m)/2}
e^{2i(\delta_l-\delta_m)}.
$$
\end{widetext}

\end{document}